\documentclass[manuscript]{aastex}

\usepackage{multirow}
\usepackage[multiple]{footmisc}
\usepackage{gensymb}

\slugcomment{Not to appear in Nonlearned J., 45.}

\shorttitle{}
\shortauthors{}

\begin{document}

\title{Chemical Composition of Young Stars in the Leading Arm of the Magellanic System\footnotemark[0]}

\author{Lan Zhang\altaffilmark{1,2,3},  Christian Moni Bidin\altaffilmark{4}, Dana I. Casetti-Dinescu\altaffilmark{5},  R\'{e}ne A. M\'{e}ndez\altaffilmark{3}, Terrence M. Girard\altaffilmark{7}, Vladimir I. Korchagin\altaffilmark{8}, Katherine Vieira\altaffilmark{9}, William F. van Altena\altaffilmark{6} \& Gang Zhao\altaffilmark{1}}

 \altaffiltext{1}{Key Lab of Optical Astronomy, National Astronomical Observatories, CAS, 20A Datun Road, Chaoyang District, 100012 Beijing, China}
 \altaffiltext{2}{CAS South America Center for Astronomy, Camino El observatorio \#1515, Las Condes, Santiago, Chile}
 \altaffiltext{3}{Departamento de Astronomia Universidad de Chile, Camino El observatorio \#1515, Las Condes, Santiago, Chile}
 \altaffiltext{4}{Instituto de Astronom\'{i}a, Universidad Cat\'{o}lica del Norte, Av. Angomos 0610, Antofagasta, Chile} 
 \altaffiltext{5}{Department of Physics, Southern Connecticut State University, 501 Crescent St., New Haven, CT 06515, USA}
 \altaffiltext{6}{Astronomy Department, Yale University, 260 Whitney Ave. , New Haven, CT 06511, USA}
 \altaffiltext{7}{14 Dunn Rd, Hamden, Connecticut, CT 06518, USA}
 \altaffiltext{8}{Institute of Physics, Southern Federal University, Stachki st/194, 344090, Rostov-on-Don, Russia}
 \altaffiltext{9}{Centro de Investigaciones de Astronomi\'{a}, Apartado Postal 264, M\'{e}rida 5101-A, Venezuela}
 \footnotetext[0]{Based on observations with the 6.5m Clay telescope at Las Campanas Observatory, Chile (program ID: CN2014A-057)}

\begin{abstract}
Chemical abundances of eight O- and B-type stars are determined from high-resolution spectra obtained with the MIKE instrument on the Magellan 6.5m Clay telescope. The sample is selected from 42 candidates of membership in the Leading Arm of the Magellanic System. Stellar parameters are measured by two independent grids of model atmospheres and analysis procedures, confirming the consistency of the stellar parameter results. Abundances of seven elements (He, C, N, O, Mg, Si, and S) are determined for the stars, as are their radial velocities and estimates of distances and ages.

Among the seven B-type stars analyzed, the five that have radial velocities compatible with membership to the LA have an average [Mg/H] of $-0.42\pm0.16$, significantly lower than the average of the remaining two [Mg/H] = $-0.07\pm0.06$ that are kinematical members of the Galactic disk. Among the five LA members, four have individual [Mg/H] abundance compatible with that in the LMC. Within errors, we can not exclude the possibility that one of these stars has a [Mg/H] consistent with the more metal-poor, SMC-like material. The remaining fifth star has a [Mg/H] close to MW values. Distances to the LA members indicate that they are at the edge of the Galactic disk, while ages are of the order of $\sim 50-70$ Myr, lower than the dynamical age of the LA, suggesting a single star-forming episode in the LA. V$_{\rm LSR}$ the LA members decreases with decreasing Magellanic longitude, confirming the results of previous LA gas studies.
\end{abstract}


\section{Introduction}
\label{sec:intro}
The Magellanic Stream \citep[MS,][]{math74, dong15}, including the Bridge \citep{kerr57, hind63}, and the Leading Arm (LA) is a well-known, $\sim 200^{\circ}$-long \ion{H}{1} structure, evidencing the interaction between the Small and Large Magellanic Clouds (SMC and LMC, respectively) and the Milky Way (MW) \citep{nide10}. In recent years, comprehensive studies of these structures have been carried out to understand the complex interactions between the Clouds and the MW, and hence to provide constraints on the gravitational potentials of these galaxies. 

The LA has a complex structure, consisting of as many as four substructures according to \citet{for13} and \citet{venz12}, situated above and below the Galactic plane and encompassing $60^{\circ}$ width. Absolute proper-motion measurements of the Clouds include the HST-based determination by \citet{kall06, kall13} and the ground-based study of \citet{viei10}.  These were used by \citet{diaz12} to explore a tidal model that simulates the Clouds' interactions. In their modeling, a leading arm can be produced from the tidal interaction, but the observed multi-branch morphology of the LA and its kinematics are not reproduced. It has been thus suggested that the LA is hydrodynamically interacting with both the Galactic gaseous disk \citep{mccl08} and the hot halo \citep{diaz11}. If this is the case, newly formed stars in the LA may be expected. 

In order to explore this hypothesis, \citet[][hereafter CD12]{case12} searched for candidate OB-type stars  over a $\sim$7900 deg$^2$ region including the Magellanic Clouds, the Bridge, the Leading Arm, and part of the Magellanic Stream. That study was based on photometry of the Galaxy Evolution Explorer Survey \citep[GALEX,][]{bian11}, the Two Micron All Sky Survey \citep[2MASS,][]{skru06}, the Southern Proper Motion Program's catalog 4 \citep[SPM4,][]{gira11}, and the American Association of Variable Star Observers All Sky Photometric Survey \citep[APASS,][]{hend11}, and on proper motions from SPM4. The outcome was a list of 567 young OB-type candidates as possible members of the MS.

In a subsequent study, \citet[][hereafter CD14]{case14} analyzed the kinematics and spectral properties of 42 of the above-mentioned OB candidates using intermediate-resolution spectra. These candidates were selected to be in three stellar overdensities in the LA region, above and below the plane. CD14 found a total of 19 young, OB-type stars. Of these, five young B-type stars had radial velocities (RVs) compatible with LA kinematics, and a low velocity dispersion of $\sim$ 33 km~s$^{-1}$.  They also found one O6V star with an age of $1 - 2$ Myr situated $\sim$ 39~kpc from the Galactic center. These discoveries suggested that recent star formation occurred in the LA as a likely consequence of  the hydrodynamical interaction between the MS and the Galactic disk where the MS crosses the Galactic plane. 

The aim of the present work is to further study the detailed chemical abundances of these young stars. We wish to 1) understand the origin of these stars and thus further constrain the history of the LA,  and 2) explore whether there is a chemical difference between young stars that are kinematical members of the LA and those that are not. To this end, we have selected a sample of eight stars: three above the Galactic plane and five below (including the O6V star). Also, of these eight stars, five are the aforesaid likely kinematical members of the LA, while three are not. We determine the abundances of elements Helium, Carbon, Nitrogen, Oxygen, Magnesium, Silicon, and Sulfur using high-resolution spectra obtained with the Magellan Inamori Kyocera Echelle \citep[MIKE;][]{bern03} on the Magellan 6.5m Clay telescope at Las Companas Observatory (Chile).

The observations are described in \S~\ref{sec:obs}. The method and the procedures of the determination of stellar parameters and abundance analysis are described in \S~\ref{sec:sp} and \S~\ref{sec:aa}. The results are presented and discussed in \S~\ref{sec:results}, \S~\ref{sec:ism}, and \S~\ref{sec:disc}, with a summary in $\S $~\ref{sec:sum}.

\section{Observations \& Data Reduction}
\label{sec:obs}

\subsection{Target selection}
\label{subsec:star_sele}
\citet{mccl08} derived a kinematic distance to the LA of $d \sim 21$~kpc from the Sun. Also, \ion{H}{1} gas velocities in the LA are in excess of 100 km~s$^{-1}$ \citep[see e.g., Fig.~8 in][]{nide10}. Therefore, we primarily selected stars whose distances are close to 21~kpc and RV $> 100$~km~s$^{-1}$ from the pilot, intermediated-resolution spectral study of CD14. In our sample, we also included three stars with RV $< 100$~km~s$^{-1}$ for the purpose of comparison. For observational reasons, faint stars ($V > 16$) were excluded. This leaves eight stars as targets in our study. We designate our stars as ``CD14-A**'' for those below the Galactic plane, at Magellanic coordinates ($\Lambda_M$, $B_M$) $\sim$ ($15^{\circ}$, $-22^{\circ}$), and as ``CD14-B**'' for those above the plane, at ($\Lambda_M$, $B_M$) $\sim$ ($42^{\circ}$, $-8^{\circ}$). The detailed spatial distribution of the stars can be seen in Fig.~1 of CD14, and we do not repeat it here. In our sample, CD14-A08 is the previously identified very hot, massive ($\sim$40~$M_{\odot}$), young ($1-2$ Myr) star with spectral type of O6V at a heliocentric distance of $\sim$39~kpc (CD14). The remaining seven stars are B-type stars. 

We list our stars in Tab.~\ref{tab:obs_log}, which includes the current designation, the SPM4 identification number, equatorial coordinates, $V$ magnitudes, as well as other observation details. 

\subsection{High-resolution spectroscopy and data reduction}
\label{subsec:reduction}
High-resolution spectra were obtained with the MIKE instrument on the 6.5m Clay telescope in March of 2014 for these eight stars. The setup gave a resolution of $R \sim 33000$ and $R \sim 29000$ for blue ($3200 - 5000$~{\AA}) and red ($4900 - 10000$~{\AA}) sides, respectively. The average seeing and airmass during the observations were 1''.2 and 1.3, respectively. The spectra were binned during the data collection. MIKE binned $3\times2$ were used, where 3 and 2 are spatial and spectral direction, respectively. Table~\ref{tab:obs_log} summarizes the observation details and exposure times.

\begin{figure}
\centering
\includegraphics[width=\textwidth]{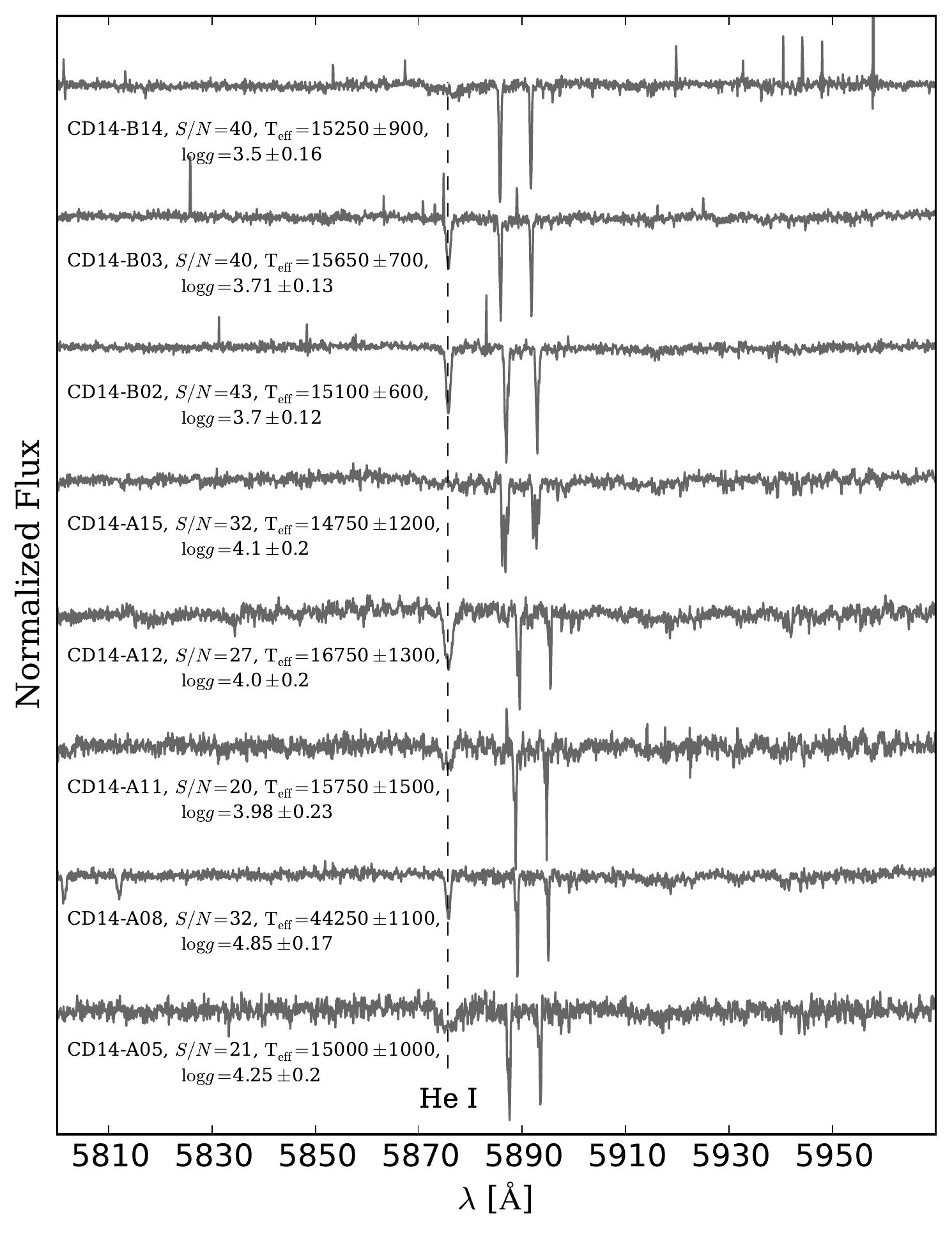}
\caption{Examples of spectra obtained with the Magellan/MIKE for the eight target stars. The vertical dashed lines indicates the position of \ion{He}{1} (5876~{\AA}).}
\label{fig:sample}
\end{figure}

The standard Carnegie Python Distribution (CarPy) routines for MIKE were used for data reduction, including order identification, wavelength calibration, flat-field correction, background subtraction, one-dimensional spectra extraction and flux calibration. The radial velocities (RV) of targets were measured by cross-correlation with a synthetic template whose temperature and gravity is similar to those of the targets after continuum rectification. We used the IRAF\footnote{IRAF is distributed by the National Optical Astronomy Observatories, which are operated by the Association of Universities for Research in Astronomy, Inc., under cooperative agreement with the National Science Foundation. \citep{tody86,tody93}} task FXCOR\footnote{\url{http://iraf.noao.edu/projects/rv/fxcor.html}} based on the standard algorithm of \citet{tonr79}, where the synthetic template was taken from the synthetic spectra library of \citet{muna05}. Two telluric oxygen A bands: the band 6884~{\AA}, transition (0,1) and the band 7621~{\AA}, transition (0,0) were also adopted to correct the RV zero point, where the molecular data were taken from the HIgh-resolution TRANsmission molecular absorption (HITRAN) database\footnote{\url{https://www.cfa.harvard.edu/hitran/}}. The signal-to-noise ratios ($S/N$s) of the spectra at 4000 and 5800 {\AA} is also presented in Tab.~\ref{tab:obs_log}. Examples of a portion of spectra are shown in Fig.~\ref{fig:sample} and the obtained RVs are listed in Tab.~\ref{tab:rv_dis_age}.

\section{Stellar Atmosphere Parameters}
\label{sec:sp}

\subsection{Measurements}
\label{subsec:meas}
The effective temperature ($T_{\rm eff}$), surface gravity ($\log g$), and atmospheric helium abundance of the target stars were measured by fitting the observed hydrogen and helium lines with synthetic spectra. The stellar model atmospheres we employed for target stars are interpolated from comprehensive grids of metal line-blanketed, non local thermodynamic equilibrium (NLTE), plane-parallel, hydrostatic model atmospheres of O-type \citep[][OSTAR2002]{lanz03a} and B-type \citep[][BSTAR2006]{lanz07} stars, with opacity sampling which is a simple Monte Carlo-like sampling of the superline cross sections and solar abundance. Both sets of grids were generated by TLUSTY, a program for calculating plane-parallel, horizontally homogeneous model stellar atmospheres in radiative and hydrostatic equilibrium \citep{hube95}.

The OSTAR2002 grid contains 12 values of $T_{\rm eff}$, $27,500~{\rm K} < T_{\rm eff} < 55,000~{\rm K}$ with 2500~K steps, eight $\log g$, $3.0 < \log g < 4.75$ with 0.25~cm~s$^{-2}$ steps, and 10 chemical compositions, from metal-rich relative to the Sun to metal-free, while the  BSTAR2006 includes 16 values of $T_{\rm eff}$, $15,000~{\rm K} < T_{\rm eff} < 30,000~{\rm K}$ with 1000~K steps, 13 $\log g$, $1.75 < \log g < 4.75$ with 0.25~cm~s$^{-2}$ steps, six chemical compositions and a microturbulent velocity of 2~km~s$^{-1}$, from twice to one-tenth of the solar metallicity and metal-free. For more details on the grids and computations of model atmospheres, we refer the reader to \citet{lanz03a, lanz07} and references therein. For all our targets, the solar-metallicity ($Z/Z_{\odot}=1$) grid is adopted. However, for CD14-A05 and CD14-A12, which show strong helium lines, an additional super-solar metallicity ($Z/Z_{\odot}=2$) grid was also employed.

The spectrum synthesis program we adopted for line synthesis is the SYNSPEC, a general spectrum synthesis program (developed by Ivan Hubeny \& Thierry Lanz) \footnote{\url{http://nova.astro.umd.edu/Synspec49/synspec.html}}. It can compute line formation under both LTE and NLTE conditions. In our analysis, the input model atmospheres are NLTE, therefore, the automatic NLTE treatment mode of SYNSPEC is employed, in which the program automatically decides which levels are treated in NLTE, and assigns proper NLTE populations to the lower and upper level of a given transition. The input model atoms and ions \citep{lanz03b} are provided by TLUSTY\footnote{\url{http://nova.astro.umd.edu/Tlusty2002/tlusty-frames-data.html}}. The form of the intrinsic line profiles is a Voigt function, in which natural, Stark, van der Waals, and thermal Doppler broadening are all included, while a Gaussian function is considered for rotational broadening. Stellar parameters are determined using a spectrum synthesis procedure that includes the Balmer series lines from H$_\alpha$ to H$_{10}$ (excepting H$_\epsilon$, which is blended with a Ca II H line) and nine \ion{He}{1} lines, i.e., 4009~{\AA}, 4026~{\AA}, 4120~{\AA}, 4143~{\AA}, 4388~{\AA}, 4471~{\AA}, 4922~{\AA}, 5876~{\AA}, and 7065~{\AA}. For CD14-A08, \ion{He}{2} (4686 ~{\AA}) is also considered because the line is sensitive to $T_{\rm eff}$.

\begin{figure}
\centering
\includegraphics[width=\textwidth]{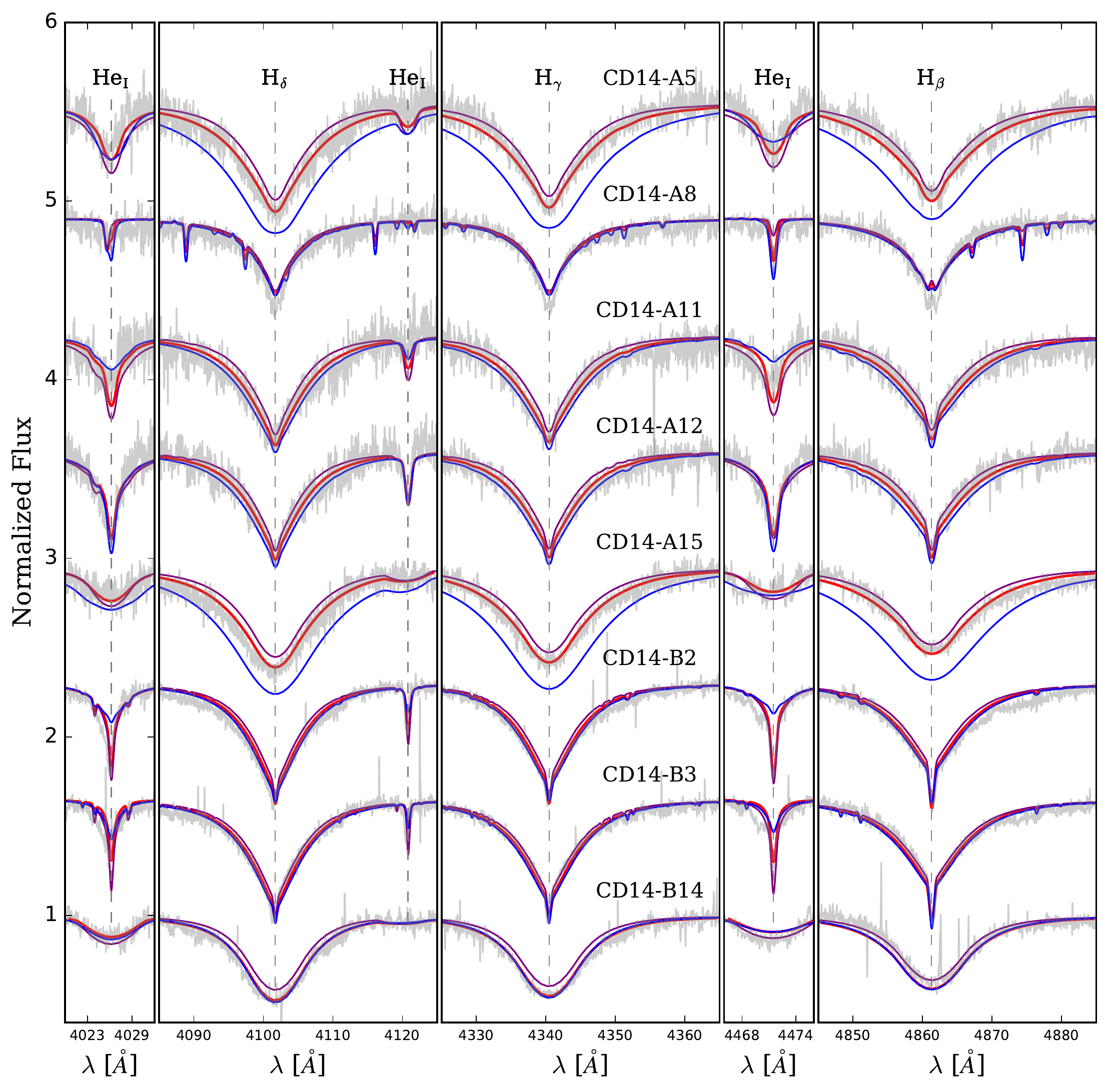}
\caption{Examples of fits to the most prominent spectral features using best-fit stellar atmosphere parameters (SA~I). Grey and red thick lines represent observed and  the best synthetic spectra, respectively. Blue and purple lines represent the synthetic spectra with $T_{\rm eff}^{\rm best} \pm \Delta T_{\rm eff}$, respectively, where $\Delta T_{\rm eff} = 2000$~K for CD14-A8, CD14-A12, and CD14-B14, while $\Delta T_{\rm eff} = 1500$~K for the rest stars. It can be seen that \ion{He}{1} lines are more sensitive to $T_{\rm eff}$.}
\label{fig:s_para_teff}
\end{figure}

\begin{figure}
\centering
\includegraphics[width=\textwidth]{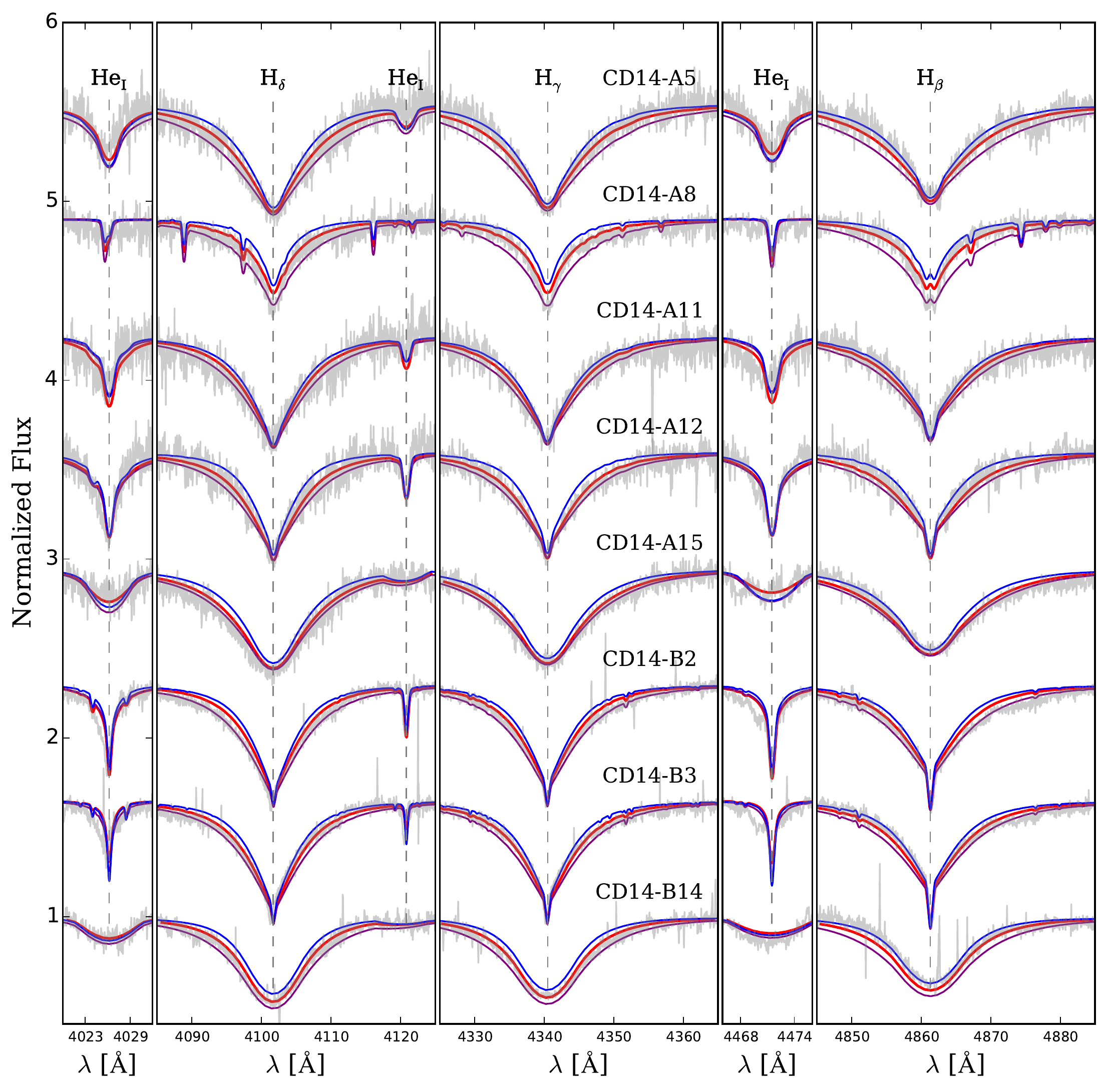}
\caption{Similar to Fig.~\ref{fig:s_para_teff}, but for variations of $\log g$. Blue and purple lines represent the synthetic spectra with $\log g_{\rm best} \pm 0.2$~dex, respectively. In this case, Balmer series lines are more sensitive to $\log g$ values.}
\label{fig:s_para_logg}
\end{figure}

The best fitted stellar parameters are found by $\chi^2$ test. n Fig.~\ref{fig:s_para_teff} the sensitivity to $T_{\rm eff}$ is shown by displaying synthetic spectra at $\pm \Delta T_{\rm eff}$ from the best-fit value of $T_{\rm eff}$. $\Delta T_{\rm eff} = 2000$~K for CD14-A08, CD14-A12, and CD14-B14. While for the rest stars, $T_{\rm eff}^{\rm best} - 2000$~K is too low to generate \ion{He}{1} profiles correctly, therefore, in this case, $\Delta T_{\rm eff} = 1500$~K.
In Fig.~\ref{fig:s_para_logg} the sensitivity to $\log g$ is illustrated by showing spectra at $\log g$ values $\pm 0.2$~dex from its best-fit value. It can be seen that $T_{\rm eff}$ and $\log g$ values are sensitive to \ion{He}{1} lines and Balmer series lines, respectively. We first used the same method as the one described in \citet[][hearafter MB12]{moni12} to estimate the errors on $T_{\rm eff}$, $\log g$ and $v\sin i$. Specifically, the uncertainties estimated from the $\chi^2$ statistics (labeled as $\sigma_{\chi^2}$) were multiplied by three to obtain the final error, since the errors propagated from the data reduction procedure, such as sky subtraction and the normalization, can't be neglected. The details are described in MB12 and references therein. But the errors of $T_{\rm eff}$ and $\log g$ are also affected by the noise level of spectra. In order to take this into account, the uncertainties propagated from the $S/N$s (labeled as $\sigma_{S/N}$) were estimated by resampling the data \citep{andr10}, which is a Monte-Carlo method. For each data point $F_{\lambda}$, we resample a new $F'_{\lambda}$ from a Gaussian distribution with a mean value of $F_{\lambda}$ and a standard deviation $\sigma_{\lambda}$, which is calculated from $\sigma_{\lambda} = \frac{F_{\lambda}}{S/N}$. Then $T_{\rm eff}$ ($\log g$) of the best fit for the new resampled spectrum is re-determined with fixing $\log g$ ($T_{\rm eff}$), $\log \frac{N_{\rm He}}{N_{\rm H}}$ and $v\sin i$. The resampling and fitting processes are repeated 50 times for each star. Thus, a series of $T_{\rm eff}$ ($\log g$) for each star is obtained. Therefore, $\sigma_{S/N}$ is given by the standard deviation of the series. The uncertainties with typical $S/N$ values are listed in Tab.~\ref{tab:err_sn}. In conclusion, the final errors on $T_{\rm eff}$ and $\log g$ are estimated by summing $3\sigma_{\chi^2}$ and $\sigma_{S/N}$ in quadrature. For the error of $\log \frac{N_{\rm He}}{N_{\rm H}}$, we refer the reader to \S~\ref{subsec:au}.  Finally, the stellar parameter results and their errors are listed in Tab.~\ref{tab:st_params}.

\subsection{Comparisons with a MB12-like analysis}
\label{subsec:com_sp}
In order to test the reliability of the stellar-parameter results, We also perform a separate, independent measuring process  using different grids of model atmospheres and analysis code (specifically, that in MB12) for the stellar parameters. In the work of MB12, the grid of model spectra is computed with Lemke's version\footnote{\url{http://a400.sternwarte.uni-erlangen.de/~ai26/linfit/linfor.html}} of the LINFOR program (developed originally by Holweger, Steffen, and Steenbock at Kiel University), which is based on local thermodynamic equilibrium (LTE) model atmospheres of ATLAS9 \citep{kuru93}. The grid covers 7000~K$~<T_{\rm eff}<~$35000~K, $2.5 < \log g < 6.0$, and $-3.0 < \log \frac{N_{\rm He}}{N_{\rm H}} < -1.0$. For CD14-A08, whose $T_{\rm eff}$ is higher than 40000~K (from the intermediate-resolution spectral study of CD14), a grid of metal-free NLTE model atmospheres of \citet{moeh04} is employed. In this analysis,  Balmer series from H$_\beta$ to H12, four \ion{He}{1} lines (4026~{\AA}, 4388~{\AA}, 4471~{\AA}, 4922~{\AA}), and two \ion{He}{2} lines (4542 {\AA} and 4686{\AA}) were fitted simultaneously. For more details on the grids of model atmospheres and analysis code, we refer the read to MB12 and references therein. 

\begin{figure}
\centering
\includegraphics[width=\textwidth]{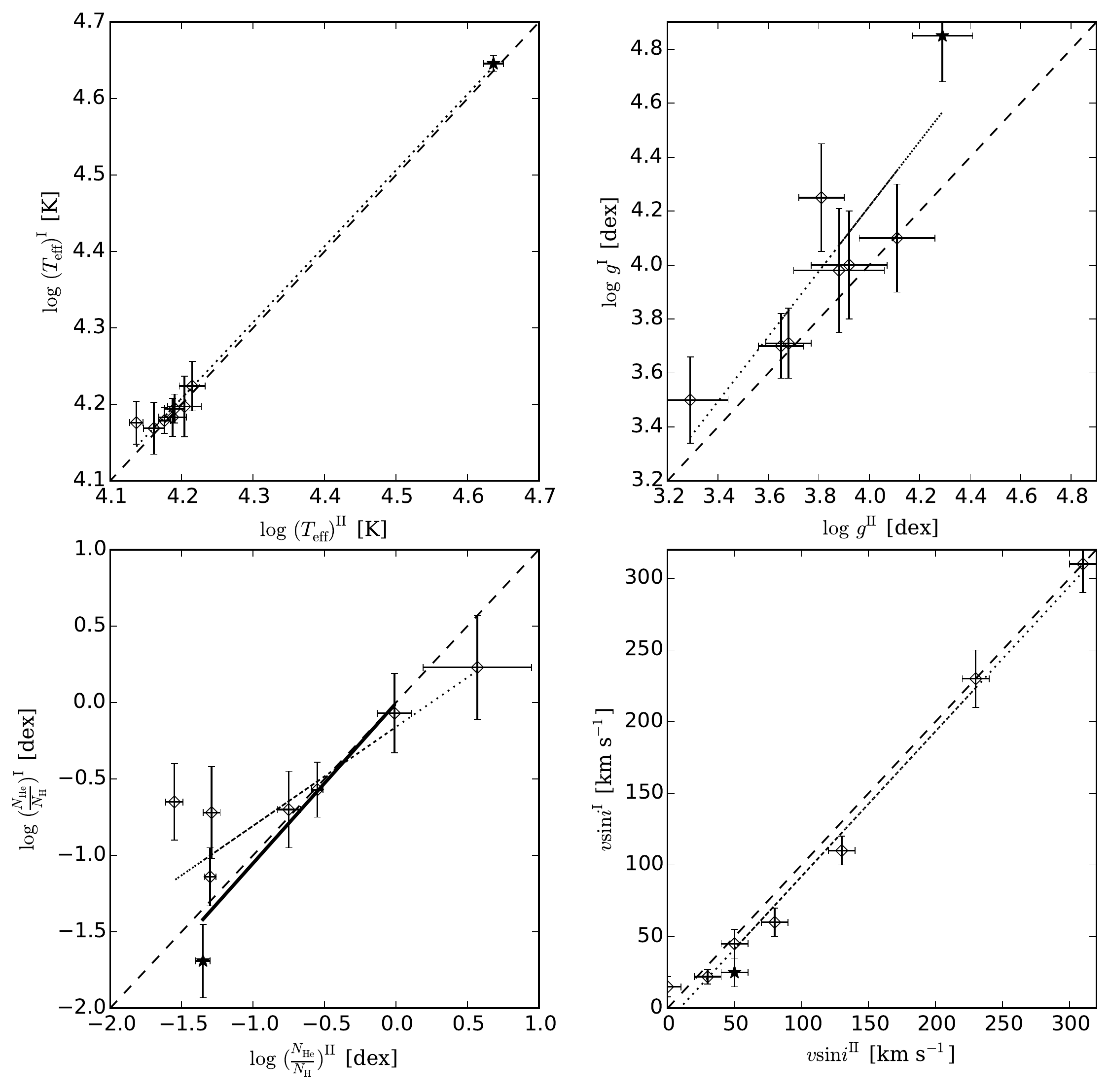}
\caption{Comparison between the two sets of stellar parameters derived from different grids of model atmospheres.  The parameters labeled ``I''  are measured using NLTE grids of TLUSTY (our default method), while those labeled ``II'' are measured with the separate, independent grid (LTE) and analysis code of LINFOR (as in MB12). The black star symbol represents our O-type target, CD14-A08. The dashed lines represent a one-to-one correlation, while the dotted line represents a linear fit to the data. In the bottom left plot, the thick solid line shows a linear fit after removing three presumed outliers.}
\label{fig:sp_com}
\end{figure}

Comparisons of the atmospheric parameters between the two analyses are presented in Tab.~\ref{tab:sp_com} and Fig.~\ref{fig:sp_com}. For convenience, we label our default spectral analysis process described in the previous subsection as ``SA~I'', and the separate one {\it \'{a} la} MB12 as ``SA~II''.

The comparison reveals a general good agreement between the two parameter sets, apart from two problematic cases. Specifically, the nature of the initially suspected hot O-type star CD14-A08 is still obscure (see Sect.~\ref{subsubsec:a8}), and the measurements for this star were challenging, while the SA~II temperature of CD14-A05 is lower than the cooler end of the BSTARS grid, hence a direct comparison of the results is doubtful. With the exclusion of these two stars, the mean difference is only 74~K in $T_{\rm eff}$ and $-0.08$~dex in $\log g$. The offset in $\log g$ seems systematic, although negligible compared to errors. SA~II $\log g$ estimates tend to be lower than SA~I values by $\approx -0.1$~dex. Rotational velocities are in good agreement (mean difference 8~km~s$^{-1}$), but it must be reminded to the reader that SA~II estimates are relatively rough, because $v\sin i$ in this method is an input value of the routine and not a fit parameter. The helium abundance shows no systematic offset between the two parameter sets, but the differences are very large in some cases. However, the SA~II results may not be very reliable, because the upper end of the employed grid is at solar abundance ($\log \frac{N_{\rm He}}{N_{\rm H}}=-1$), while most of the stars (six, according to SA~I) have super-solar values.

A star-by-star analysis reveals that the two sets of results show good agreement within the $1-\sigma$ region for CD14-A12, CD14-B02, CD14-B03, and CD14-B14, while for other stars:
\begin{itemize}
\item[-] {\it CD14-A05} - the two sets of stellar parameter results show large discrepancy, but within $1-\sigma$ range.  The main reason to cause the discrepancy is that the cooler end of BSTARS grid (used in SA~I) is 15000~K. Although the model atmospheres could be extrapolated from the grid, it is only in a case that the $T_{\rm eff}$ \& $\log g$ values are not far from the limits of the grid, that is, smaller than the grid steps. Therefore, the parameter values of SA~I are the best ones we could find in the possible grid ranges of BSTARS. On the other hand, the SA~II results are also problematic, because this star is extremely helium-rich. The SA~II routine extrapolated the helium abundance beyond the model grid by about 1.5~dex, and this fact can easily affect the resulting stellar parameters. We found that forcing a lower helium abundance in the SA~II routine could increase $\log g$ up to $4.0 - 4.2$~dex, consistent with the SA~I value, but never increase the resulting temperature beyond 13800~K. With the SA~II results, BSTARS and SYNSPEC failed to generate \ion{He}{1} 5876~{\AA}, and 7065~{\AA} lines and generated too broad Balmer series lines for the observed spectrum. As a reference, a grid of super-solar model atmosphere is adopted to estimate the stellar parameters, and the results are $T_{\rm eff} = 15200\pm1100$~K, $\log g = 4.30\pm0.20$~dex, $\log \frac{N_{\rm He}}{N_{\rm H}} = -0.00\pm 0.32$~dex, and $v\sin i =110\pm10$~km~s$^{-1}$. Although the super-solar metallicity atmospheres results in a lower He abundance, the $T_{\rm eff}$, $\log g$, and $v\sin i$ are consistent with the SA~I's results in $1-\sigma$ region. 
\item[-] {\it CD14-A08} - the $T_{\rm eff}$ and $\log g$ from SA~I are much higher than the ones of SA~II. The reasons of this discrepancy is that neither SA~I nor SA~II can fit the Balmer series lines perfectly for this star, that is, the cores and the wings of Balmer series lines can not be fit simultaneously. Within the SA~II method, the cores are weighted more, and model wings are shallower than those observed, which leads to a lower $\log g$. Since \ion{N}{3} --- which lies on the wings of H$_{\delta}$ --- can be detected and therefore used to determine the nitrogen abundance, in SA~I, the fitting process is weighted in favor of fitting the wings. The line cores are consequently poorly fit, as can be seen in Fig.~\ref{fig:a8_sample}. The final $\log g$ value falls outside the grid of both SA~I and SA~II, because it is higher than the upper limits of OSTARS and lower than the lower limit of the \citet{moeh04} grid. However, in the case of SA~I, the out-of-grid 0.1~dex extrapolation is smaller than the grid step of 0.25~dex.
\item[-] {\it CD14-A11 \& CD14-A15} - for these two stars, the $T_{\rm eff}$, $\log g$, and $v\sin i$ of SA~I and SA~II are consistent within $1-\sigma$. But the helium abundances of SA~I are higher than those of SA~II by 0.57 and 0.90~dex, respectively. For CD14-A11, SA~II derived a 250~K higher $T_{\rm eff}$, which leads to a $\sim 0.10$~dex lower He abundance,  a 0.10~dex lower $\log g$ and a 20~km~s$^{-1}$ lower $v\sin i$, which results in a $\sim 0.18$~dex higher He abundance. Therefore, a higher $\log \frac{N_{\rm He}}{N_{\rm H}}$ value from SA~II is expected. The situation of CD14-A15 is quite similar. We explore possible reasons for this discrepancy by making the following tests for the two stars:
\begin{itemize}
 \item[i.] we used routines of SA~I with the $T_{\rm eff}$, $\log g$ and $v \sin i$ derived from SA~II to re-measure helium abundances, and we obtained higher values, i.e., $\log \frac{N_{\rm He}}{N_{\rm H}} = -0.60\pm 0.25$~dex  and $\log \frac{N_{\rm He}}{N_{\rm H}} = -0.57\pm0.22$~dex for CD14-A11 and CD14-A15, respectively. The results are in good agreement with the values from SA~I.
 \item[ii.] we repeated the SA~II process, fixing helium abundances to values derived from SA~I then redetermining $T_{\rm eff}$, $\log g$ and $v\sin i$, and found that the differences between new $T_{\rm eff}^{\rm SA~II}$, $\log g^{\rm SA~II}$ and $v\sin i^{\rm SA~II}$ and previous SA~II ones are within $1-\sigma$. 
\end{itemize}
The tests confirm that $T_{\rm eff}$, $\log g$ and $v\sin i$ are only slightly affected by the helium abundance, and that the helium abundance derived with SA~II is somehow underestimated.
\end{itemize}

For the parameters $T_{\rm eff}$, $\log g$ and $v\sin i$ which are mainly measured from the Balmer series lines, the two sets of stellar atmospheric parameters agree reasonably with each other. Removing three outliers, namely, CD14-A05 whose model atmosphere is limited by the present grid of BSTARS and CD14-A11 \& CD14-A15 whose $\log \frac{N_{\rm He}}{N_{\rm H}}$ values were obviously underestimated by SA~II, the two sets of He abundance results are also in a good agreement. As discussed above, with SA~II parameter results, SYNSPEC can't generate correct \ion{He}{1} lines and Balmer line wings for CD14-A05 and CD14-A08, respectively. For consistency, we will adopt SA~I results for following abundance analysis.

\subsection{Age and distance calculation}
\begin{figure}
\centering
\includegraphics[width=\textwidth]{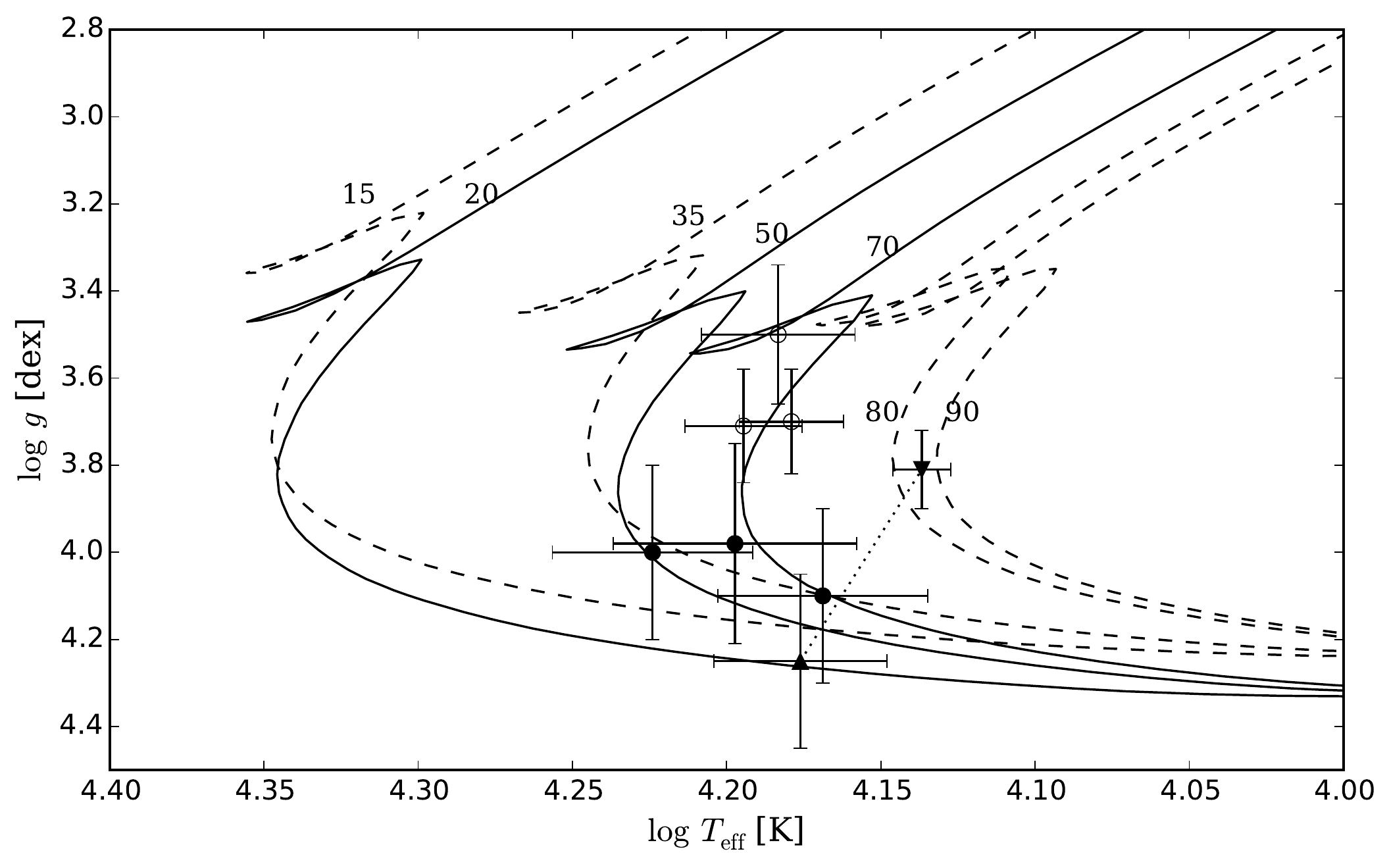}
\caption{The location of target stars with respect to solar-metallicity (black solid lines) and super-solar metallicity (black dashed lines) PARSEC isochrones \citep{bres12}. Ages (in Myr) of each isochrone are labeled. Open and filled symbols represent stars that lie above and below the Galactic plane, respectively. The up and down triangles which connected with a dotted line represent the location of CD14-A05 with $T_{\rm eff}$ and $\log g$ from SA~I and SA~II, respectively. }
\label{fig:isochrone}
\end{figure}

With the measured stellar parameters, we determine the age of the target stars via comparison of their position in the temperature-gravity plane with solar-metallicity PARSEC isochrones \citep{bres12}. The uncertainty in this estimate is given by the error box associated with the stellar parameters. This is shown in Fig.~\ref{fig:isochrone}. The isochrones also return an estimate of the absolute magnitude $M_V$ of each star. Thus, we calculate the true distance moduli $(m-M)_0$ from the apparent $V$ magnitudes of the SPM4 catalog, de-reddened with the \citet{schlegel98} maps as corrected by \citet{boni00}, and assuming a standard reddening law with  $R_V = \frac{A_V}{E(B-V)}=3.1$. We have verified that the results are not affected within errors by the use of the \citet{schlafly11} calibration, and that the distance moduli obtained using the 2MASS $J$ and $K_s$ magnitudes are consistent with those obtained in $V$, with no systematics. We have also tested our results with an alternative method, estimating the spectral type of each star from its temperature and gravity, and then using the spectral type-absolute magnitude calibration of \citet{wegn06}. The star CD14-A08 was excluded from this determination, because its nature (a true main-sequence or a sdO star, see Sect~\ref{subsubsec:a8}) is uncertain, and the procedure could lead to misleading conclusions if this star is misinterpreted.

\section{Abundance Analysis}
\label{sec:aa}
Using the stellar parameters obtained with the stellar model atmospheres and synthesis code described in \S~\ref{sec:sp},  we proceeded to determine the element abundance of Carbon, Nitrogen, Oxygen, Neon, Magnesium, Silicon, and Sulfur by fitting observed lines with synthetic spectra. We did not determine the iron abundance because there are no clear, unambiguous iron lines that can be detected in the current low $S/N$. 

\subsection{Line synthesis}
\label{subsec:ls}
\begin{figure}
\centering
\includegraphics[width=\textwidth]{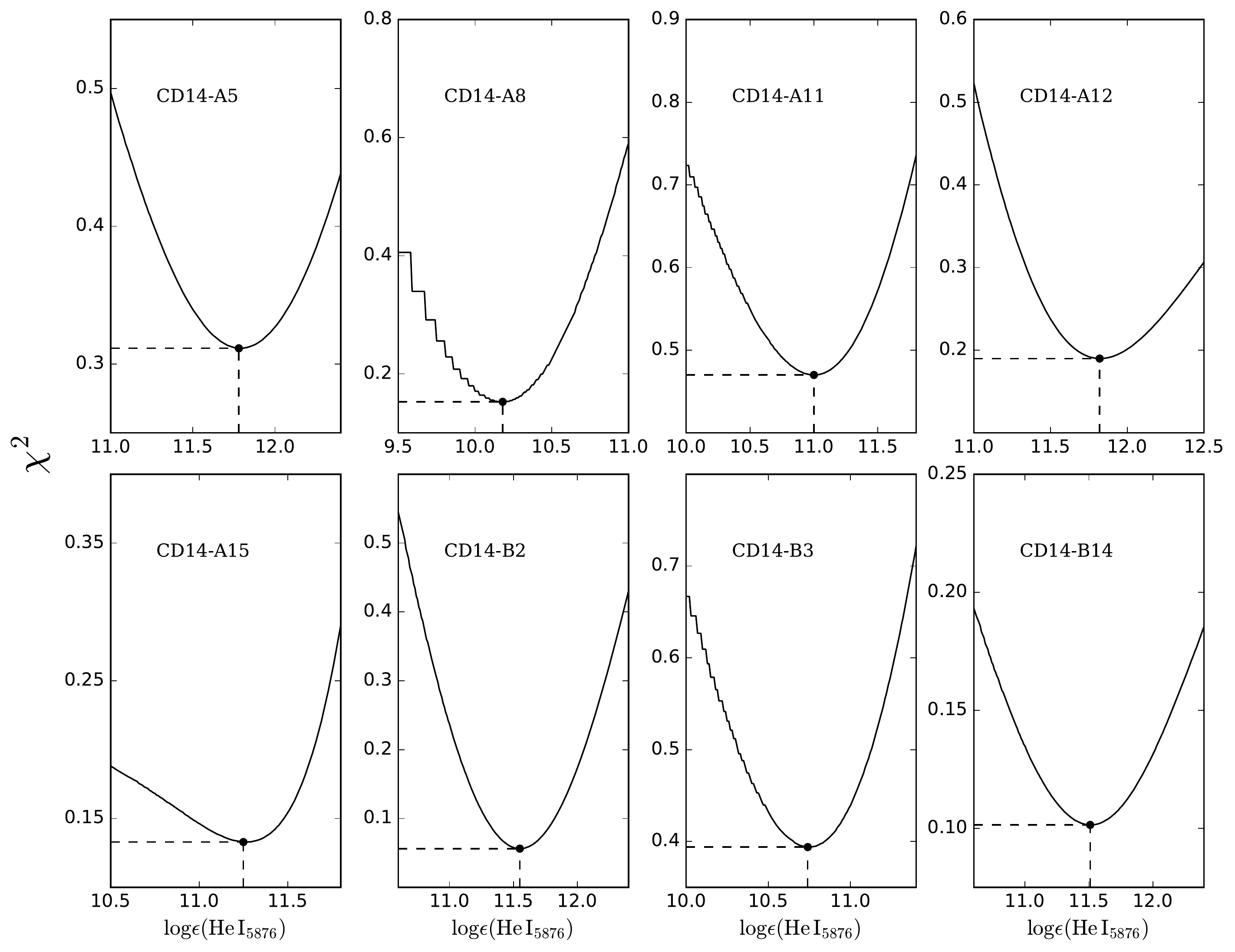}
\caption{$\chi^2$ value as a function of abundance. \ion{He}{1}~$\lambda$~5876 is shown as an example.}
\label{fig:chi2_example}
\end{figure}

We adopt the solar composition from \citet{grev98} in this process. The atomic line data of elements, including $\log gf$ values and central wavelength are mainly selected from \citet[][K95]{kuru95}. For some lines, the absorption oscillator strength $\log gf$ is selected from the NIST Atomic Spectra Database\footnote{\url{http://www.nist.gov/pml/data/asd.cfm}}. If the value on NIST is labeled with ``B''  or higher grade, which means that the uncertainty region of $\log gf$ is $< 10\%$, the NIST value was used in our analysis instead of that from K95. For each detected unblended line, the best fitted abundance value is also found by $\chi^2$ test. Fig.~\ref{fig:chi2_example} shows an example of finding the best abundance values. For blended lines -- for instance, the \ion{S}{2} triplet, which is blended with \ion{He}{1} at 4142 to 4148 {\AA} -- we varied Sulfur abundance with fixed Helium abundance to derive the best fit for the whole spectrum in the wavelength range. The mean abundance of each element is derived from all lines that can be detected. For stars in which feature lines for certain elements cannot be detected clearly, the feature that is at the position of the theoretical absorption line was fitted, and the maximum value for this element that could fit the spectrum is considered as the upper limit for its abundance. The synthetic spectra fitted to the observed ones with its corresponding $1-\sigma$ error range are shown in Fig.~\ref{fig:a8_sample} $-$ Fig.~\ref{fig:mg_sample}, using representative \ion{C}{2}, \ion{N}{2}, \ion{N}{3}, \ion{O}{2}, \ion{O}{3}, \ion{Mg}{2} 4481 {\AA}, \ion{Si}{2}, and \ion{S}{2} as examples.

\subsection{Abundance determination and uncertainties}
\label{subsec:au}
For each element, synthetic spectra are produced using the stellar parameters measured with the SA I procedure and with various abundance values. These synthetic spectra are then compared with the observed ones. A $\chi^2$ test was employed to establish the best fit for detected absorption lines.  Uncertainties in abundances are mainly caused by (1) observation error; (2) uncertainties in the analysis of individual lines, including random errors of atomic data and fitting uncertainties; (3) errors in the continuum rectification; (4) uncertainties of the stellar parameters; (5) statistical error from line-to-line analysis.

\begin{figure}
\centering
\includegraphics[width=\textwidth]{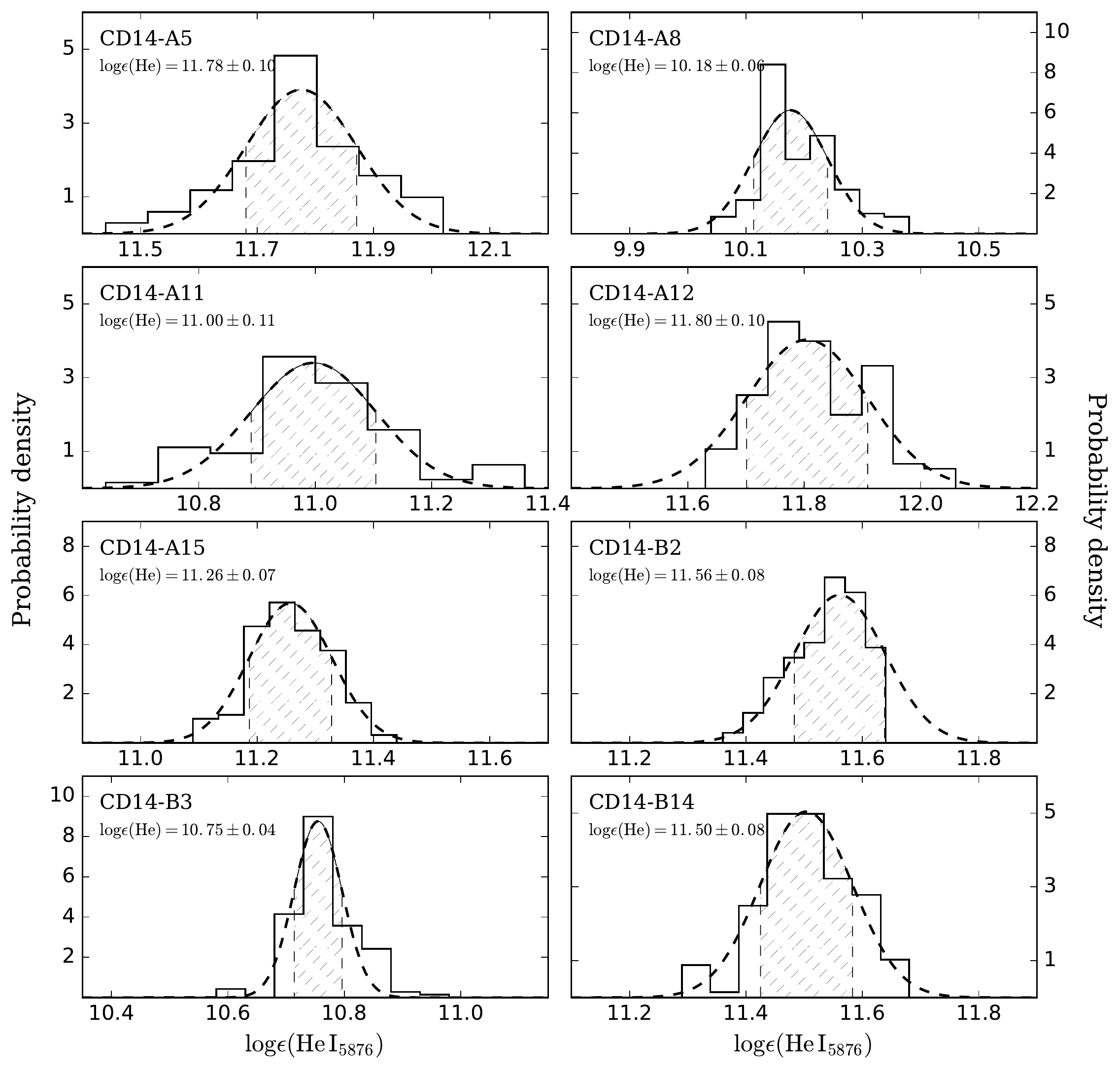}
\caption{The abundance uncertainties from the observation error and the fitting process. \ion{He}{1} $\lambda$~5876 is shown as an example. Histograms are distributions of $\log \epsilon$(\ion{He}{1}$_{5876}$) for sample stars. Dashed lines and shades are the best fits for the distributions and $1-\sigma$ uncertainty regions, respectively.}
\label{fig:err_pdf}
\end{figure}

The uncertainties from the observation error are also estimated by resampling the data (see \S~\ref{subsec:meas}). The resampling and abundance determination processes are repeated 50 times for each star. Thus, we obtain a distribution of abundance values for each star. In our analysis, the distributions are approximately Gaussian. Therefore, the best abundance values and their uncertainties are given by the mean and the standard deviation of the Monte-Carlo distributions.  An example of how we derive the uncertainties is shown in Fig.~\ref{fig:err_pdf}. The uncertainties with typical $S/N$s are listed in the last row of Tab.~\ref{tab:err_sn}. 

An uncertainty of 10\% in $\log gf$ was also adopted to explore its effect on the abundance. This results in an error of 0.02 dex on average. 

The continuum around some lines, such as \ion{He}{1} 3965 {\AA}, \ion{C}{2} 6578 and 6583 {\AA}, is affected by the wings of the Balmer series lines. It is thus difficult to pinpoint an accurate continuum location for this wavelength range, which has a direct effect on the abundance determination. In the worst case, the error in continuum rectification was estimated to be 5\%, which results in a change of the abundance of up to 0.05~dex.

Errors in the stellar parameters listed in Tab.~\ref{tab:st_params} will propagate into the final abundance determinations. The effect on the abundances is estimated by varying atmospheric parameters and rotational velocity as listed in Tab.~\ref{tab:err_at_pro} for two stars in our sample: CD14-A08 (O-type star) and CD14-B02 (B-type star).

The scatter in the abundance determinations from different lines gives another estimate of the uncertainty. This error is estimated by dividing the standard deviation ($\sigma$) of the derived abundances from individual lines by the square root of the number of lines used ($N^{\frac{1}{2}}$).  For instance, the standard deviation of the abundance determinations from \ion{He}{1} lines of CD14-B02 is 0.14~dex, and the uncertainty ($\sigma \cdot N^{\frac{1}{2}}$) is about 0.05~dex. For elements with only a small number of lines available (e.g., C, O), the $\sigma$ of \ion{He}{1} (typically 0.07~dex) is adopted instead of the $\sigma$ of those species.

Finally, the overall abundance uncertainty is estimated by summing the uncertainties from these five sources in quadrature.

\section{Results}
\label{sec:results}

\subsection{Abundances}
\label{subsec:abun_res}
The derived abundances are summarized in Tab.~\ref{tab:abun_results_a} and shown in Fig.~\ref{fig:abun_res} (triangles/red symbols). For CD14-A05 and CD14-A12, abundances derived from the metal-rich model atmospheres ($Z/Z_{\odot}$) are also shown for comparison (see yellow symbols of Fig.~\ref{fig:abun_res}). Abundances derived using the metal-rich model atmosphere appear slightly, but systematically, lower than those obtained using the solar metallicity model atmosphere; however, values are consistent within their $1-\sigma$ errors. 

Mg abundance of all B-type stars from our sample are subsolar or near-solar, within uncertainties. CD14-A08, the only O-type star, has super-solar Mg abundance, and will be discussed in detail below. He abundance appears super-solar in all B-type stars, except CD14-B03 which is slightly sub-solar, but within uncertainties may be solar. C abundances of the B-type stars show enrichments in different degrees above solar, except CD14-A11 which is near solar. 

We proceed now to discuss individual stellar results. First, we discuss the O-type star CD14-A08, and then the B-type stars.

\subsubsection{CD14-A08}
\label{subsubsec:a8}
\begin{figure}
\centering
\includegraphics[width=\textwidth]{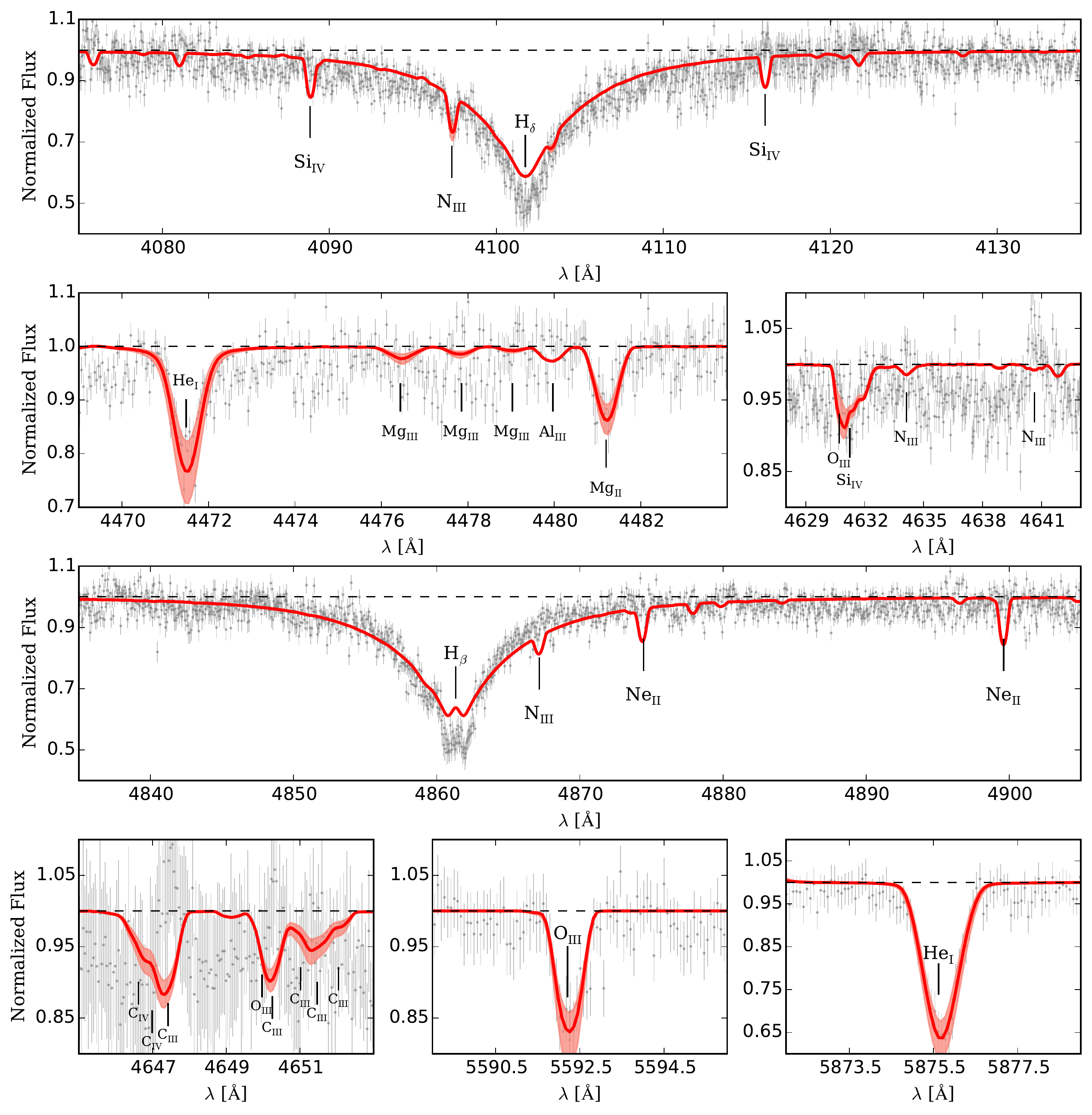}
\caption{Examples of fits of observed spectra (black dots) with synthetic spectra (thick lines) for CD14-A08. The shaded area shows the $1-\sigma$ fitting error range.}
\label{fig:a8_sample}
\end{figure}

\begin{figure}
\centering
\includegraphics[width=\textwidth]{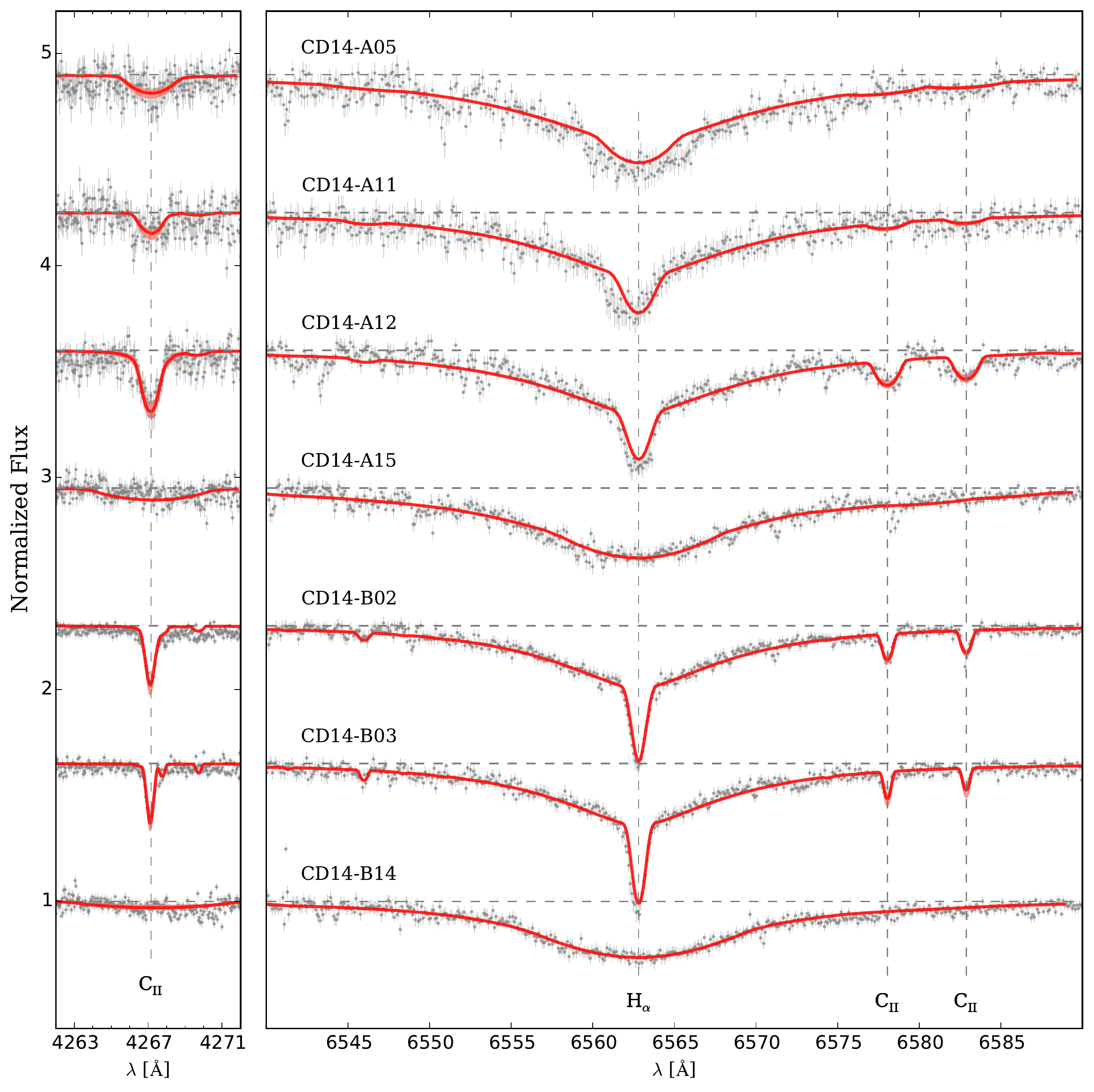}
\caption{Examples of fits of observed spectra (black dots) with synthetic spectra (thick lines) for the \ion{C}{2} lines of our B-type target stars.}
\label{fig:c_sample}
\end{figure}

\begin{figure}
\centering
\includegraphics[width=\textwidth]{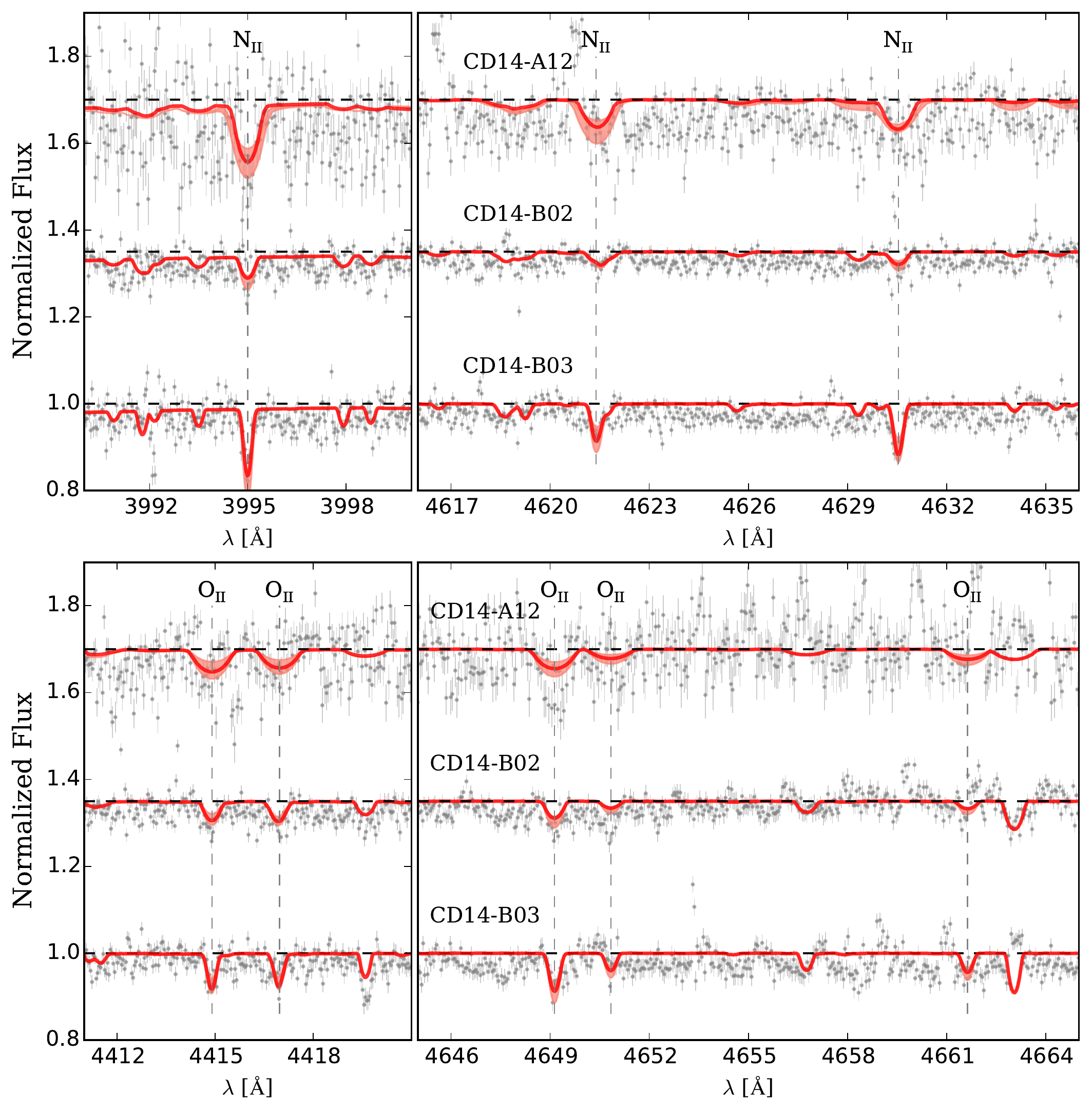}
\caption{Same as Fig.~\ref{fig:c_sample}, but for representative \ion{N}{2} and  \ion{O}{2} lines. Only stars for which  \ion{O}{2} and \ion{N}{2} lines could be detected are shown.}
\label{fig:o_sample}
\end{figure}

\begin{figure}
\centering
\includegraphics[width=\textwidth]{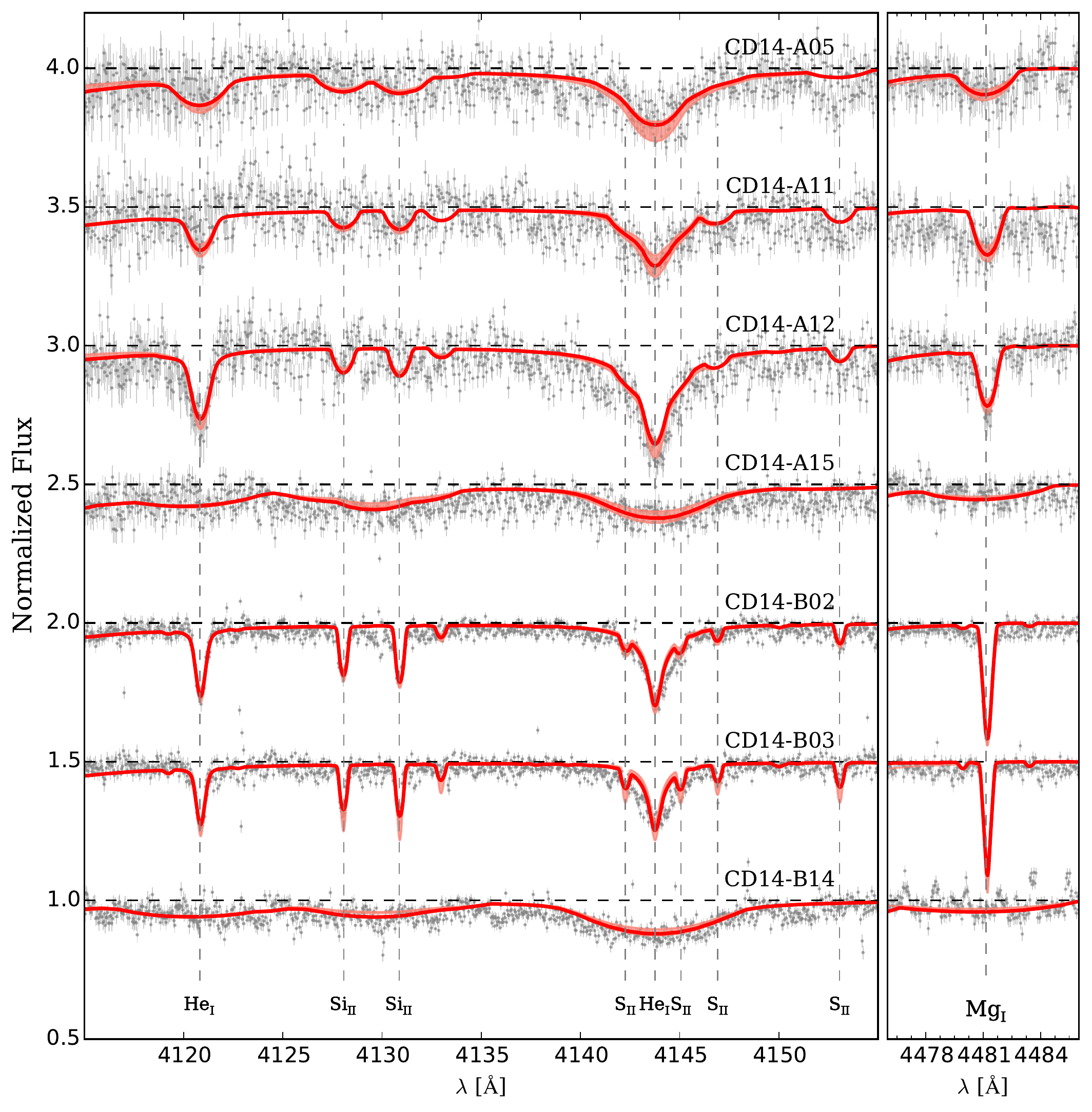}
\caption{Same as Fig.~\ref{fig:c_sample}, but for \ion{Mg}{2} 4481 {\AA}, \ion{Si}{2}, and \ion{S}{2} lines.}
\label{fig:mg_sample}
\end{figure}

\begin{figure}
\centering
\includegraphics[width=\textwidth]{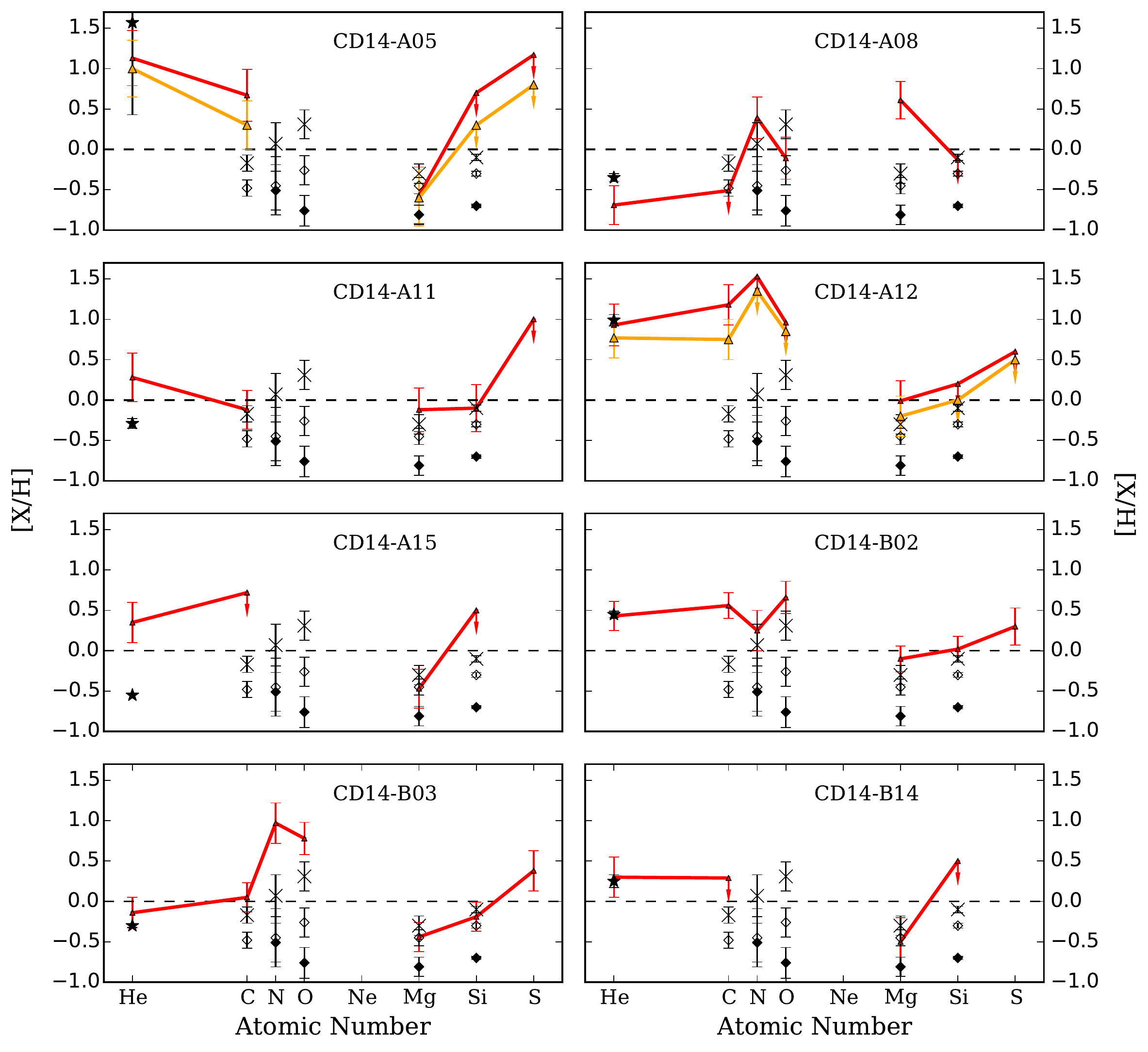}
\caption{Abundance results derived using the stellar parameters from SA I. Red symbols show our results. Yellow symbols are results derived from the super-solar metallicity grid of model atmospheres (see text). Downward arrows indicate that only upper limits of the abundances can be given. The filled star symbol represents the helium abundance determined from SA II. Weighted average abundances of N, O, Mg, and Si of B-type stars in three clusters are shown as references. Specifically, NGC 4755 (in the MW), NGC 2004 (in the LMC), and  NGC 330 (in the SMC) from \citet{trun07} are represented with crosses, open diamonds, and filled diamonds, respectively. The C abundances of NGC2004-D15 (in the LMC, open diamonds) and HR 3468 (in the solar  neighbourhood, crosses) from \citet{przy08} also are shown for reference.}
\label{fig:abun_res}
\end{figure}

This star was classified  as an O6V star in CD14. As discussed in CD14, if CD14-A08 is an O6V star, its young age ($1-2$~Myr) and large Galactocentric distance of $\sim 39$~kpc (CD14) suggest that it was born {\it in situ}. However, with its updated stellar parameters of $T_{\rm eff} = 44250\pm1100$~K, $\log g = 4.85\pm0.17$~dex, and [He/H] = $-0.69\pm0.24$~dex, which is measured by fitting the wings of Balmer series lines, CD14-A08 is more likely an under-luminous He-deficient subdwarf O (sdO) star. \citet{hebe09,hebe16} summarized abundance features of sdO stars: 1) only a relatively small fraction of sdO stars are He-poor; 2) such He-deficient sdO stars show no C and N; 3) element abundances of He-deficient sdOs are similar to those of subdwarf B (sdB) stars, in which O and Mg are on average slightly subsolar.  Besides, sdO stars are typically Pop II stars with large velocities, found at high galactic latitudes. In the spectra of CD14-A08, no clear and notable C lines can be detected because of low $S/N$, and only upper limits of C abundance can be given, which is actually much lower than solar abundance. Detectable \ion{O}{3} 5592~{\AA} gives slight sub-solar O abundance, which is also expected for sdOs. However, strong \ion{N}{3} 4097~{\AA} (see upper panel of Fig.~\ref{fig:a8_sample}) and \ion{Mg}{2} 4481~{\AA} lines were detected, implying super-solar abundance that is unusual for an sdO. With a relatively low $\log g$ for an sdO, it is possible that diffusion processes may favor an enhancement in Mg (U. Heber, private communication), but the N enhancement still cannot be explained. Based on the atmospheric parameter and abundance features, the nature of CD14-A08 is uncertain. For this hot star, abundance determinations could be modified by selective emission effects. However, according to \citet{heap06} and \citet{walb10}, metal lines which appear in either absorption or emission in the optical spectral region include \ion{N}{4}~4058~{\AA}, \ion{Si}{4}~4088~{\AA}, 4116~{\AA}, \ion{N}{3}~4634 -- 4641~{\AA}, \ion{C}{3}~4647 -- 4651~{\AA}, and \ion{S}{4}~4486~{\AA}, 4503~{\AA}. Among these line, \ion{Si}{4}~4088~{\AA}, 4116~{\AA} occurs only at very high temperatures ($T_{rm eff} > 50,000$~K), and then only weakly. If the emission effects were not considered, the abundance would be somehow overestimated. However, because of the low S/N of the spectrum, the metal lines listed above can not be detected clearly, especially for S (there are no [S/H] estimated for the star). It is difficult to explore the emission effects in detail from the present observations. For this reason, we only list upper limits of [X/H] for C, Si. As we described above, [N/H] was estimated from \ion{N}{3}~4097~{\AA}. \ion{N}{3}~4634 -- 4641~{\AA} are not clear enough to be analyzed (see the third plot of Fig.~\ref{fig:a8_sample}). For this star, we list all the information that we can obtain from the present data. Its nature remains uncertain. For this reason we didn't include the star in the following discussion about the LA. To further explore the nature of this star, more data need to be gathered and analyzed, which is beyond the purpose of this paper.

\subsubsection{CD14-A05}
\label{subsubsec:a5}
It is a He-rich star with [He/H] = $1.13\pm0.34$. Even considering its uncertainty, the helium abundance is large. This star also shows evidence of carbon enhancement. That is, intermediate \ion{C}{2} 4075 and 4267~{\AA} are detected, which give an abundance result of $\log \epsilon({\rm C})$ = $9.20\pm0.32$.  \ion{Si}{2} lines are buried in the noise of its spectra. They are not very clear, therefore, we present only an upper limit of Si abundance. Because of its rapid $v\sin i$, \ion{S}{2} lines combined with \ion{He}{1} 4143~{\AA} can not been separated clearly. Only an  upper limit of S abundance can be estimated by fitting line wings of  \ion{He}{1} 4143~{\AA}.

\subsubsection{CD14-A11}
\label{subsubsec:a11}
\ion{He}{1} lines, \ion{C}{2} 4075~{\AA} and  \ion{Mg}{2} 4481~{\AA} are broader than single lines for which rotational broadening is considered. If a larger $v\sin i$ is considered, Balmer series lines can't be fitted well. The present $v\sin i$ may lead to overestimates of the abundance for these elements during the fitting process, with consequently larger error bars.

\subsubsection{CD-A15 \& CD14-B14}
\label{subsubsec:a15}
These two stars have large rotational velocities. Weak lines cannot be detected due to the broadening effects of rotation. Only strong lines such as \ion{He}{1} 5876~{\AA},  \ion{Mg}{2} 4481~{\AA} and  \ion{C}{2} 4267~{\AA} can be detected and used in the abundance determination. The measurement of stellar parameters and determination of element abundances is made more difficult by having so few line features. Therefore, we do not have sufficient information for reliable abundance determination for these two stars. Also, rotational broadening implies that the abundance is insensitive to the variance of the stellar parameters. Therefore, the uncertainties of the abundances for the two stars are relatively large. We only show upper limits of carbon abundance for the two stars although super-solar carbon abundance is suggested by the broad, flat \ion{C}{2} 4267~{\AA} line. 

\subsubsection{CD14-A12, CD14-B02, \& CD14-B03}
\label{subsubsec:a12}
Most of the absorption lines used in our chemical analysis are detected for these three stars. The average $S/N$ values for CD14-B02 and CD14-B03 are the highest in our stellar sample. Similar to CD14-A05, CD14-A12 is a helium-rich star with strong carbon enhancement even when the super-solar metallicity model atmosphere is employed. Strong \ion{N}{2} 3995~{\AA} can be detected in the spectra of CD14-A12 and CD14-B03, which represents a significant nitrogen enhancement. Abundances of Mg and Si for these three stars are all slightly sub-solar, while S abundance is super solar.

\subsection{Radial velocities}
\label{subsec:rv_res}
The results of our RV measurements are given in Tab.~\ref{tab:rv_dis_age}. Three stars below the Galactic plane (CD14-A08, CD14-A11, CD14-A12) show a RV lower than 75~km~s$^{-1}$, compatible within $2-\sigma$ with the kinematics of thin+thick disk populations derived from the Besancon Galactic model \citep{robi03}, as shown in Fig.~\ref{fig:rv_abun}. These stars are likely not LA members. However, CD14-A11 and CD14-A12 are at $z\approx2$ and $\approx3$~kpc from the plane, respectively. Young stars can be found far from the nominal $b=0$ plane at some Galactic longitudes due to the warped and flared thin disk \citep[e.g.,][]{carr15}, but this is not expected here in the fourth Galactic quadrant. Hence, these two stars most probably did not form in situ, but are Galactic runaway stars leaving their birthplace in the Galactic plane at high velocity. The remaining five stars have RVs in excess of 130~km~s$^{-1}$, and consistent with LA membership.

A comparison between our RV measurements and the first-epoch values of CD14 shows excellent agreement. The RV differences are taken as ours$-$CD14, hereafter, with errors being the quadratic sum of the uncertainties of the two measurements. The star CD14-A08 shows the largest RV variation ($\Delta RV=-23$~km~s$^{-1}$) but this is still not overly significant ($1.7\sigma$). The nature of this star is also unclear (see Sect.~\ref{subsubsec:a8}), so it will not be considered in the discussion. All the other stars show variations lower than $\vert \Delta RV \vert=5$~km~s$^{-1}$, which is nearly half the average errors, and very tiny compared to the huge variations typical of massive B-type binary stars. The only exception is the target CD14-A12, whose variation of $\Delta RV=17$~km~s$^{-1}$ is significant only at $1.7\sigma$.

The binary fraction $f$ of Galactic massive stars is very high, with preference for close systems with large RV variations. For example, \citet{kouw07} found $f>$70\% at a 3-$\sigma$ confidence level among A- and B- stars in the Scorpio OB2 association. Thus, our null detection is surprising, because the probability of having no binary system among seven stars is lower than 1\% if $f>$50\%. However, a null detection does not mean the absence of binaries, as the result must be weighted by the detection efficiency of the observations. To explore the significance of our result, we modeled a population of binaries as in \citet{moni06}, and we checked the fraction of detections given our two-epochs observations and a detection threshold of $\Delta RV=20$km~s$^{-1}$. The synthetic population was uniformly distributed in $v\sin i$, phase, and mass ratio, with a $M=5~M_\odot$ primary, and a uniform distribution of periods in the range $P$=0--100d \citep[][where the large majority of massive binary stars are found]{sana08}. We thus found that the mean detection probability of our observations, weighted on the period distribution, is $\overline{d}$=0.49. Inserting this value in Equation~4 of \citet{moni09}, we find that the probability of our null detection out of seven stars is 14\% (8.7\%) if the underlying binary fraction is 50\% (60\%). We conclude that our result implies no compelling statistical evidence that the binary fraction in our sample is lower than the normal Galactic population.

\subsection{Distances and ages}
Our estimates of ages and distance moduli for the B-type target stars are given in Tab.~\ref{tab:rv_dis_age}. We used super-solar metallicity isochrones with [Fe/H]=+0.5 for CD14-A12, due to the high $\log \frac{N_{\rm He}}{N_{\rm H}}$ derived in Sect.~\ref{sec:sp}. The star CD14-A05 would also require the same treatment, but its gravity ($\log g = 4.25\pm0.22$, SA~I) is too high for a metal-rich object, and no isochrone matches its position in the diagram. A proper estimate is impossible in this case. Considering the $1-\sigma$ $\log g$ error, the lower limit $\log g = 4.03$ provides an upper limit to the distance modulus of $(m-M)_0<15.7$. This points to an extremely young age ($<$12~Myr). Therefore, we will use a solar metallicity isochrone to estimate its age and distance. It results in a distance modulus of $15.4\pm0.3$ and an age of $25\pm15$~Myr. It can be seen that the upper limit value derived from the metal-rich isochrones falls in the $1-\sigma$ region. Besides, this is the only star for which SA~I and SA~II methods returned notably different stellar parameters (see Tab.~\ref{tab:st_params} and Sec.~\ref{subsec:com_sp}). The SA~II parameters return a larger distance, and an age compatible with the other stars of the sample, as shown in Tab.~\ref{tab:rv_dis_age}.

\section{ISM Absorption}
\label{sec:ism}
We also investigate interstellar medium (ISM) absorption along the line-of-sight of our target stars, in order to check that whether the star is behind MS matter or not. Absorption features of different ISM components  (\ion{Ca}{2}~HK lines, \ion{Na}{1}~D lines, and \ion{K}{1} line) are shown in Fig.~\ref{fig:ism}, plotted on the radial-velocity scale. For stars above the Galactic plane (i.e., CD14-B**), the absorption feature is simple, with RVs of different lines being very close to one another. The average RV of the five ISM absorption lines in the spectrum of CD14-B14 is $-3.60\pm10$~km~s$^{-1}$, which is comparable with the gaseouos Galactic thin disk \citep{gala15}. For the remaining two stars above the plane, CD14-B02 and CD14-B03, only red-shifted components can be found.  Specifically, the average ISM RVs for these two stars are $13\pm6$~km~s$^{-1}$ and $12\pm5$~km~s$^{-1}$, respectively.  Checking the  \ion{H}{1} profiles along each line-of-sight (see Fig. \ref{fig:h_profile}), extracted from the combined Leiden-Argentine-Bonn survey \citep[LAB,][]{Kalb05,fich89} of Galactic \ion{H}{1}\footnote{data were extracted from the LAB Survey server of the Argelander-Institut für Astronomie (AIfA), which was set up for the EU under grant 510308-LLP-1-2010-1-FR-COMENIUS-CMP.}\footnote{\url{https://www.astro.uni-bonn.de/hisurvey/euhou/LABprofile/}}, we notice that towards CD14-A** the absorption from the MW consist blue-shifted components while the remaining pointing show red-shifted components with RVs $>$ 15~km~s$^{-1}$. Considering the uncertainties in RV measurements and the resolution of the spectra, the red-shifted ISM absorptions detected in the spectra of CD14-B02 and CD14-B03 also may be from the Galaxy.

For the stars located below the Galactic plane, CD14-A**, Doppler splitting is observed along the lines-of-sight. RVs span from $-40$~km~s$^{-1}$ to $+50$~km~s$^{-1}$. From the CO survey of the MW \citep{dame01}, the lines-of-sight of these stars intersect the foreground Chamaeleon Clouds \citep{whit97, alve14}, with an average radial velocity of $-7$~km~s$^{-1}$ \citep{corr04}. Considering both the data of the LAB survey and the CO survey, the ISM absorption detected in the spectra of CD14-A** stars are from the MW and the foreground Chamaeleon Clouds.

We do not find any significant ISM absorption at RVs in excess of 100 $-$ 150~km~s$^{-1}$. 

\section{Discussion}
\label{sec:disc}
\subsection{General Trends in Abundances}
\label{subsec:general}
\citet{trun07} derived abundance patterns for 61 B-type stars in four fields centered on MW clusters NGC 3293 and NGC 4755, and on the LMC and SMC clusters NGC 2004 and NGC 330 respectively, with TLUSTY. The type of stars and the measuring procedure is close to our work, therefore rendering a meaningful comparison. In our discussion, we choose NGC 4755 since its location is closer to our targets than NGC 3293 (but NGC 3293 and NGC 4775 have similar mean elemental abundances). We investigate the abundance patterns of our target stars by comparing with the abundance patterns of the stars in NGC 4755, NGC 2004, and NGC 330, chosen as representatives of the three stellar populations: MW, LMC and SMC respectively.

For each cluster, we use the weighted average abundances of N, O, Mg and Si for B-type stars. In Fig.~\ref{fig:abun_res}, we present abundances for each target star (triangle symbols), as well as for the three clusters: crosses for NGC 4755, open diamonds for NGC 2004 and filled diamonds for NGC 330. Also, C abundances of NGC2004-D15 (in the LMC, open diamonds) and HR 3468 (in the solar neighbourhood, crosses) from \citet{przy08} are referenced. It can be seen that C, N, and O are highly enhanced in most of the stars, while the two $\alpha-$elements, Mg and Si, are not. Our measurements suggest that the Mg abundance of our target B-type stars is higher than that of stars in NGC 330, and consistent with stars in NGC 4755 and NGC 2004, at a $1-\sigma$ level.  Although the Si abundance can be determined only for three stars (others are presented as upper limits), this abundance shows a similar behavior to that of Mg. For stars whose CNO abundances can be determined, CD14-A11 shows [C/H] compatible with HR 3468, and CD14-B02 shows [N/H] and [O/H] similar to NGC 4775. Thus, our first conclusion as derived from Fig.~\ref{fig:abun_res} is that the environment in which our target stars formed is similar to that of the MW and the LMC, and unlike that of the more metal-poor SMC. Due to relatively large uncertainties in abundance determinations for CD14-A15 and CD14-B14, these stars' Mg abundance may be consistent with the values in NGC 330. Therefore, we can not exclude the possibility that some of our target stars may have formed from material as metal-poor as that in the SMC.

\subsection{Abundances versus Radial Velocity and Distance Modulus}
\label{subsec:abund_rv_dist}
\begin{figure}
\centering
\includegraphics[width=0.7\textwidth]{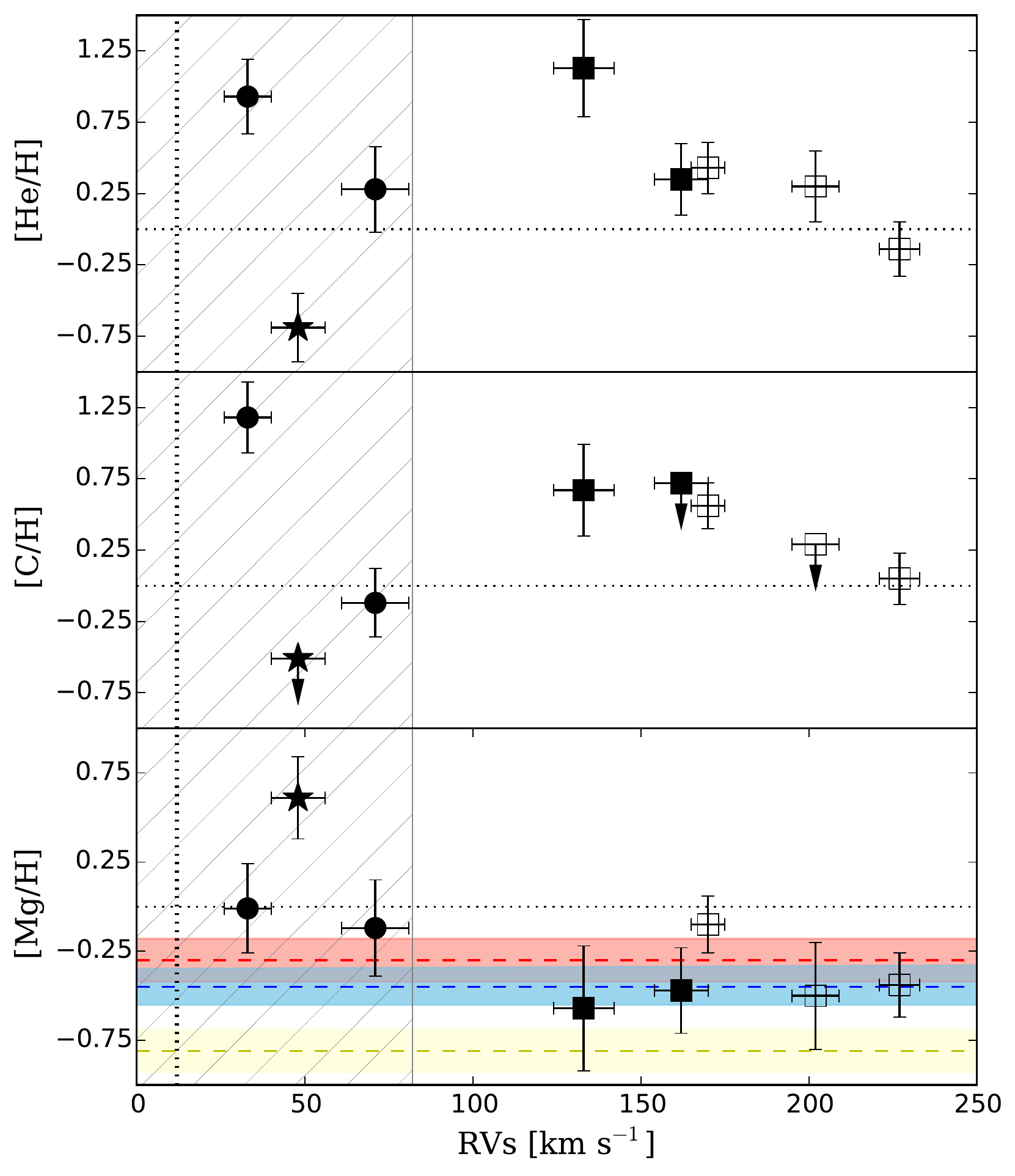}
\caption{Abundance results for He, C, and Mg as a function of RV. Open and filled symbols represent stars that lie above and below the Galactic plane, respectively. Downward arrows show upper limits of specific elements.  The possible LA and MW members (kinematically selected) are represented by squares and circles, respectively. The black star symbol is CD14-A08. The horizontal dotted lines indicate the mean solar abundance of a given element. The vertical dotted line shows the mean value for the Galactic thin+thick disk \citep{robi03}. The hatched area represents a $2-\sigma$ range about this mean. In the bottom panel, red, blue, and yellow dashed lines and their corresponding color-coded shaded areas represent the average [Mg/H] and $1-\sigma$ region of B stars in NGC 4755 (MW), NGC 2004 (LMC), and  NGC 330 (SMC), respectively.}
\label{fig:rv_abun}
\end{figure}

\begin{figure}
\centering
\includegraphics[width=0.75\textwidth]{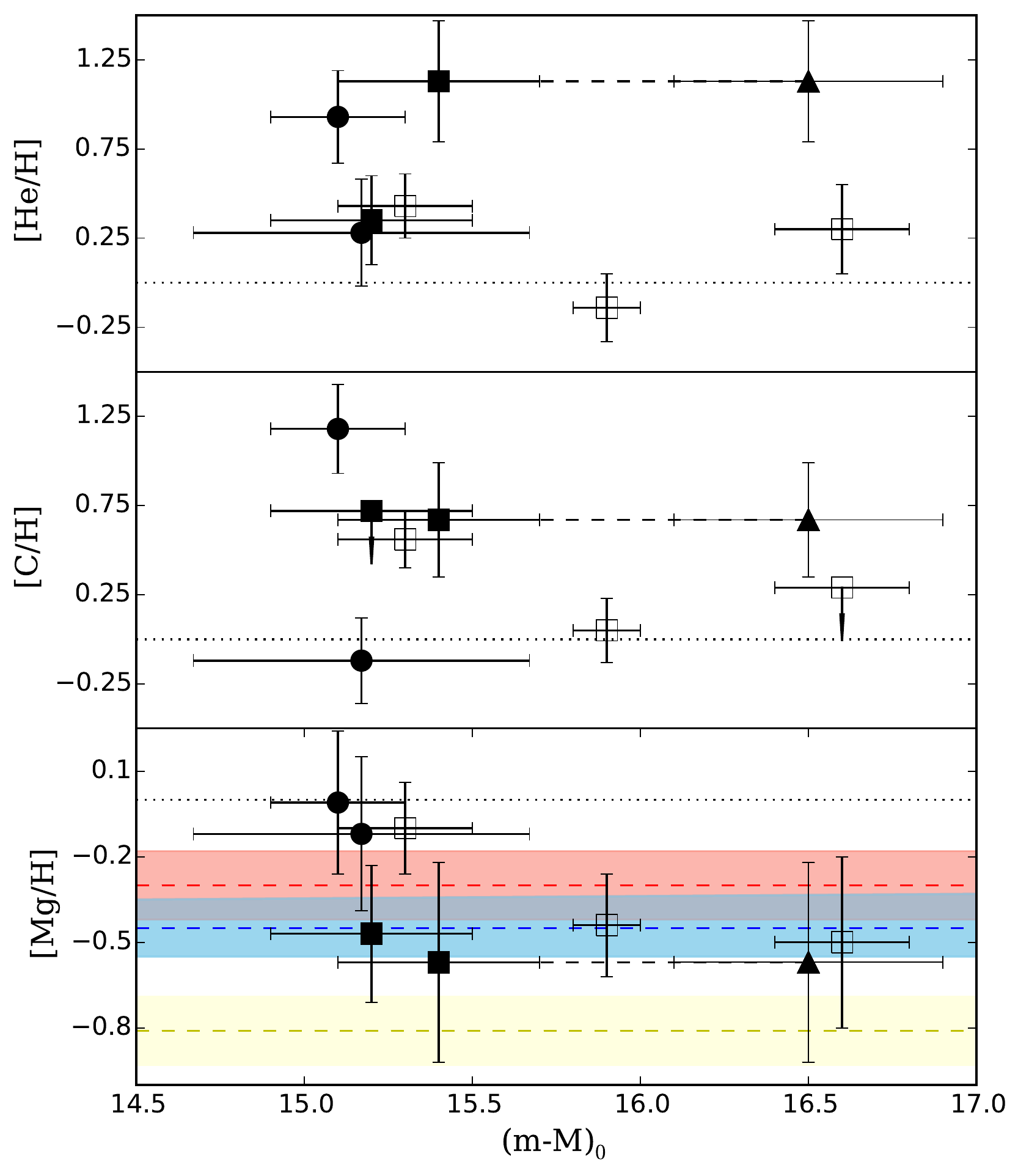}
\caption{Similar to Fig.~\ref{fig:rv_abun}, but for abundance results of B-type stars as a function of the distance modulus. For CD14-A05, we also show the distance modulus calculated with SA~II parameters as a black triangle. Its two distance modulus results are connected with a black dashed line.}
\label{fig:dis_abun}
\end{figure}

Next, we investigate abundances as a function of kinematics and distance modulus for our target stars. We inspect only He, C and Mg, since these are best determined for the entire set of stars. Figs.~\ref{fig:rv_abun} and \ref{fig:dis_abun} show the abundance ratios for He, C, and, Mg as a function of the RV and of the distance modulus, respectively. In Fig.\ref{fig:rv_abun}, the black star symbol represents CD14-A08. Filled and open symbols correspond to stars located below and above the Galactic plane, respectively. In Fig.~\ref{fig:rv_abun}, the average and $2-\sigma$ range of RV for the thin+thick disk populations is also indicated, as derived from the Besancon galactic model \citep{robi03}. In the bottom panels of Figs.~\ref{fig:rv_abun} and \ref{fig:dis_abun} we also indicate the average [Mg/H] abundance for our three representative samples; the MW (red dashed line), the LMC (blue dashed line) and the SMC (yellow dashed line). The color-coded shaded areas correspond to $1-\sigma$ ranges around the averages.

At first glance, no obvious trend of [X/H] against the RV or the distance modulus is seen in our data. However, there is a suggestion that stars with ${\rm RV} > 100$~km~s$^{-1}$ have lower Mg abundance than those with disk-like velocities. We note that the only star with a large RV ($170\pm5$~km~s$^{-1}$), and a somewhat large Mg abundance, is CD14-B02. For stars with ${\rm RV} > 100$~km~s$^{-1}$, He and C abundances tend to decrease with increasing RV and show smaller scatters than those of stars with ${\rm RV} < 100$~km~s$^{-1}$.

In what follows we will consider stars with RVs in excess of 100 km~s$^{-1}$ to be kinematical members of the LA, while the remaining stars are assumed to be kinematical members of the MW disk. We argue this as follows.
First, the RV dispersion of the five high RV stars is $\sim$ 39~km~s$^{-1}$, which is too low compared to $\sim$ 130~km~s$^{-1}$ of Galactic runaway stars \citep[][CD14]{brom09}; therefore a disk runaway origin is unlikely. 
Second, the stars are at the location of $275\degree < l < 305\degree, 10\degree < |b| < 14\degree$, where they are out of the Galactic Southern warp \citep[$180\degree < l < 270\degree$,][]{carr15}. Their average Galactocentric distance ($R_{\rm GC}$) is $\sim 13$~kpc (adopting $R_{\odot} = 8.0$~kpc). Galactic stellar disc thickness shows a rather constant scale-height within $R_{\rm GC} \sim 15$~kpc and then flares out to $R_{\rm GC} \sim 23$~kpc \citep{moma06}. In our sample, only CD14-B14 is at $R_{\rm GC} \sim 19$~kpc, but the RV of the star is 202~km~s$^{-1}$ which is much higher than the Galactic disk value. Thus, a warp origin is unlikely. Therefore, in our sample we have two stars, CD14-A11 and CD14-A12 as kinematical members of the Galactic disk/MW; and five stars, CD14-A05, CD14-A15, CD14-B02, CD14-B03, and CD14-B14 as kinematical members of the LA (see also Section ~\ref{subsec:rv_res}). Calculating the average [Mg/H] for the two groups of stars, we obtain [Mg/H] = $-0.07\pm0.07$ for the disk sample, and [Mg/H]=$-0.42\pm0.16$ for the LA sample. Clearly, the LA sample is more metal poor than the disk sample, and in agreement with the abundance of cluster NGC 2004 in the LMC ([Mg/H]=$-0.45\pm0.10$). Also, the larger [Mg/H] scatter of the LA group implies that the source of LA material is complicated.

Fig.~\ref{fig:dis_abun} displays abundances versus distance modulus. The only apparent trend is for [Mg/H] which seems to decrease with distance. Also, at large distances the [Mg/H] scatter is lower than at low distances, suggesting that the low-distance group is more inhomogeneous than the large distance one (albeit the small number of stars). We note that the sole LA kinematical member with high [Mg/H] abundance ($-0.10\pm0.16$), namely star CD14-B02, also has a small distance among the five LA members.

\subsection{A Magellanic-coordinates View of Our Sample}
\label{subsec:mag_coord}
\begin{figure}
\centering
\includegraphics[width=\textwidth]{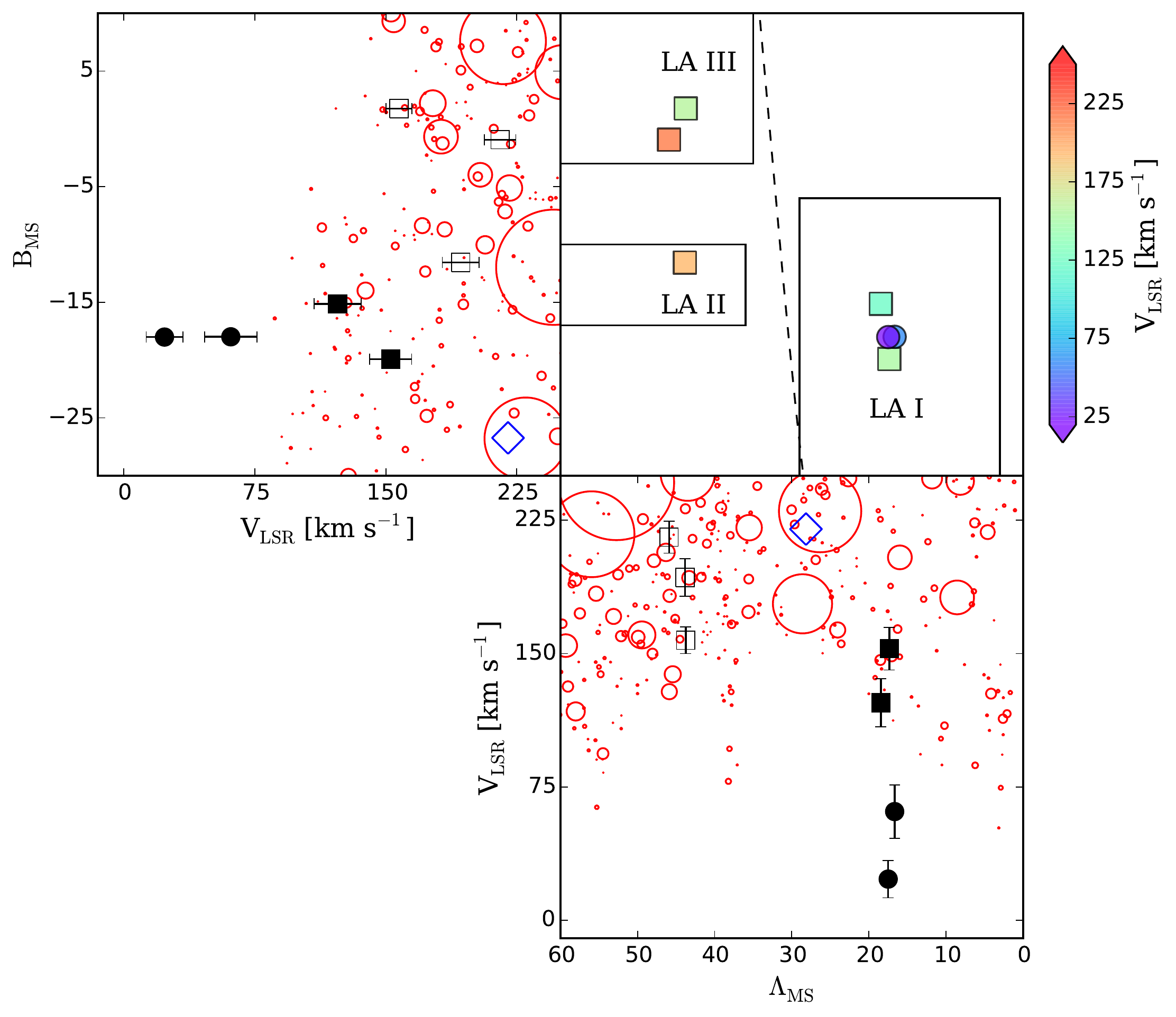}
\caption{The V$_{\rm LSR}$ distribution on the MS coordinate system. In the upper left and the bottom right plots, symbols have the same meaning as those in Fig.~\ref{fig:rv_abun}. In the upper right plot, filled squares and circles indicate the kinematically selected LA and MW members, respectively. The three different LA regions are labeled following the partitioning in \citet{nide10}. In the upper-left and bottom-right plots, blue open diamonds represent the HVC~$306-2+230$ from \citet{mccl08}, and red open circles represent the LA \ion{H}{1} clouds from \citet{venz12}. The diameter of these open circles scales with the mass of each cloud. The dashed line represents the Galactic plane.}
\label{fig:mage_rv}
\end{figure}

\begin{figure}
\centering
\includegraphics[width=\textwidth]{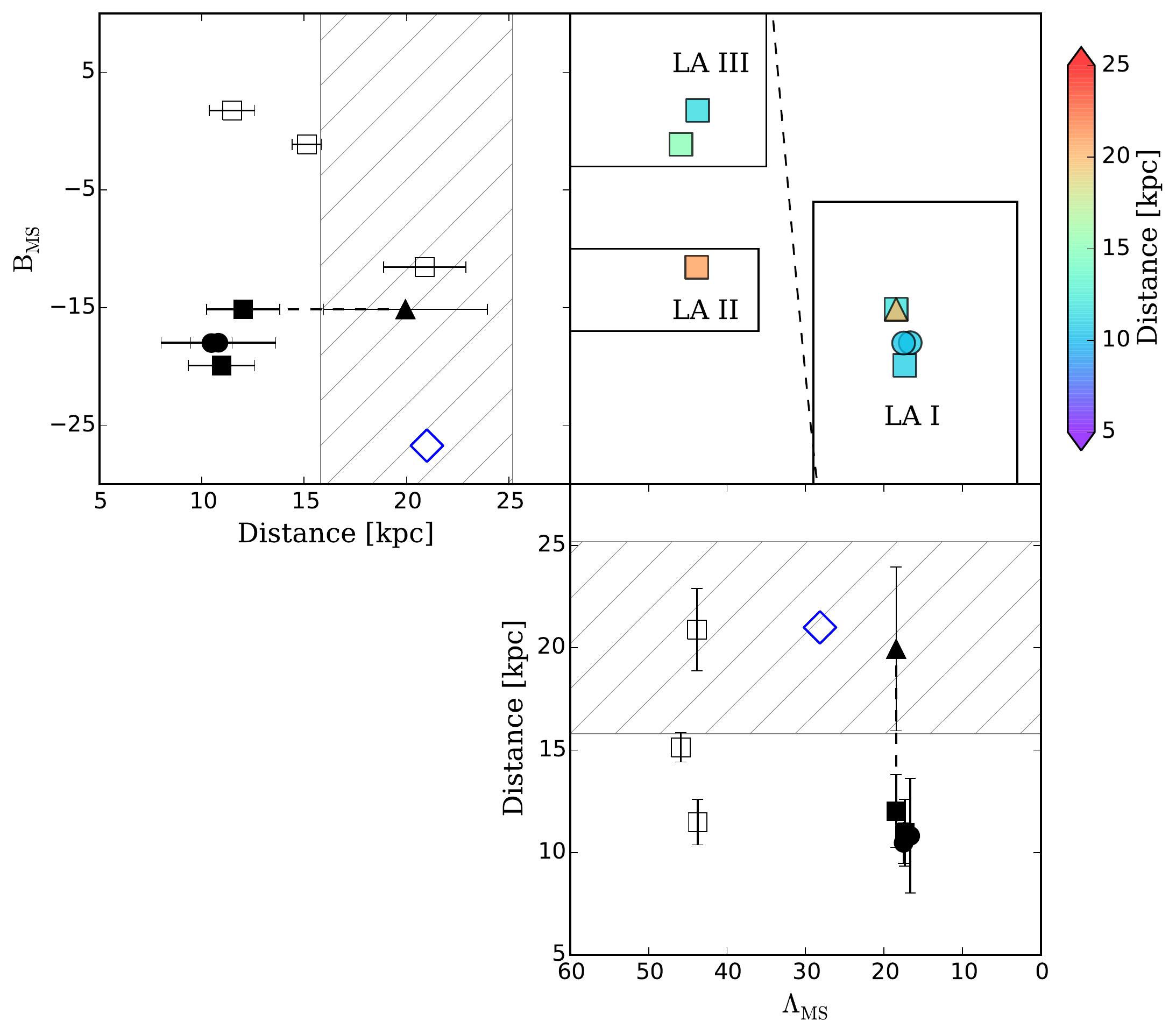}
\caption{Similar to Fig.~\ref{fig:mage_rv}. But for the distance distribution. The shades represent the kinematical distance of one high velocity cloud member of the LA \citep{mccl08}; the width of the shades corresponds to a 20\% error in the distance.}
\label{fig:mage_dist}
\end{figure}

\begin{figure}
\centering
\includegraphics[width=\textwidth]{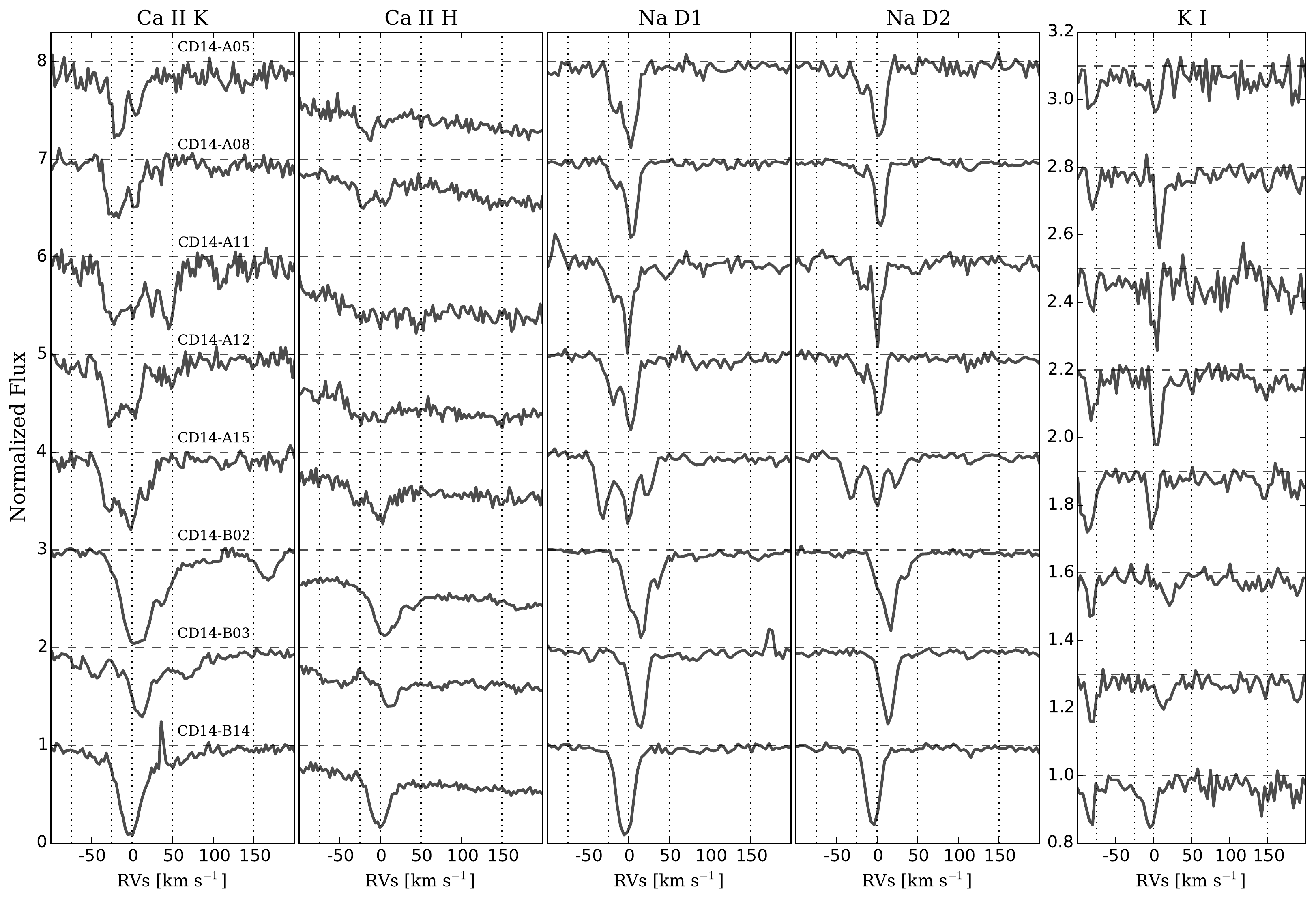}
\caption{Identified interstellar features in the spectra of our targets, plotted on the heliocentric radial-velocity scale.}
\label{fig:ism}
\end{figure}

\begin{figure}
\centering
\includegraphics[width=\textwidth]{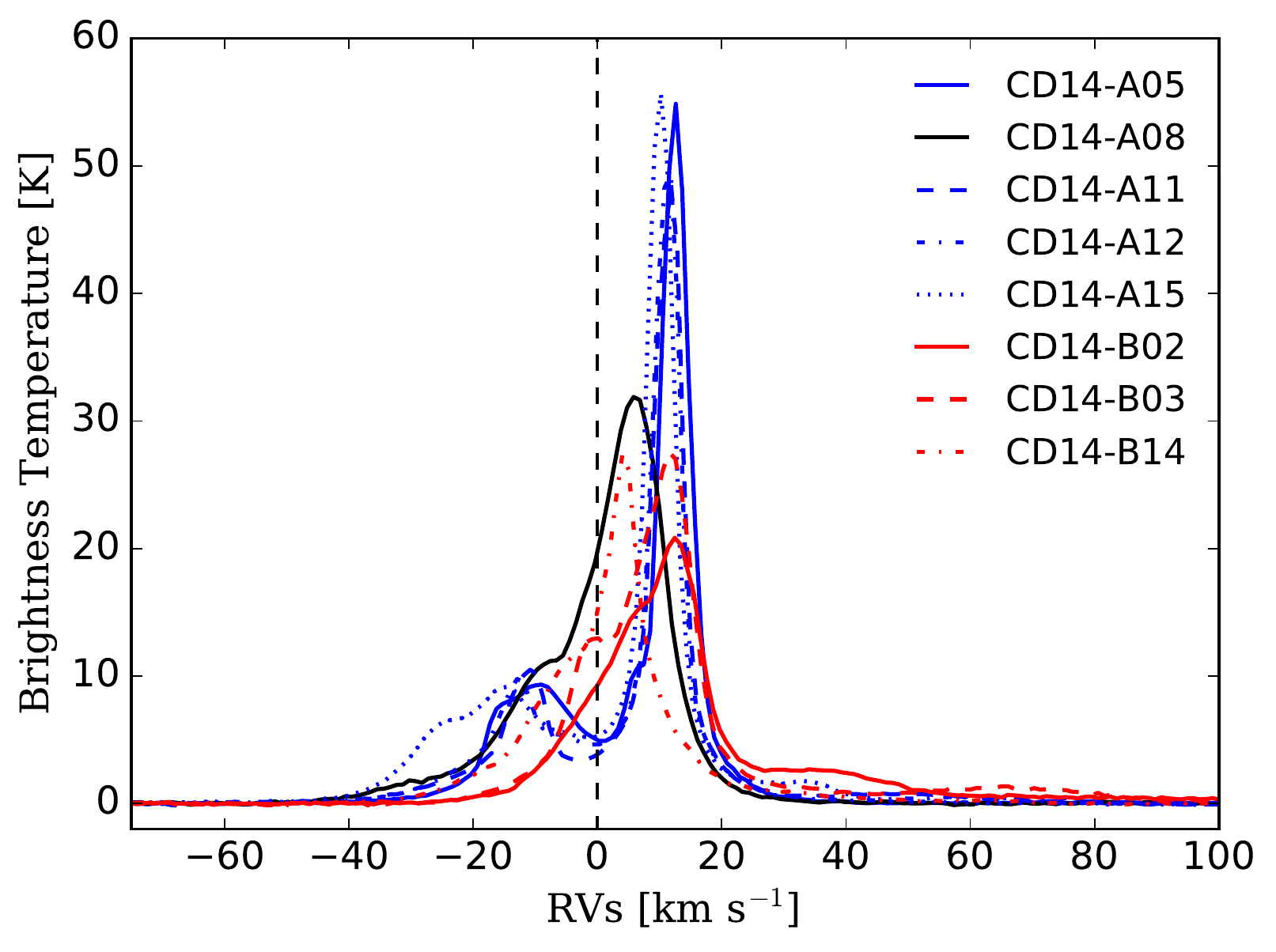}
\caption{The \ion{H}{1} profiles towards each line-of-sight. Data are extracted from the AIfA EU-HOU Survey server. The $V_{\rm LSR}$ is converted to heliocentric radial velocity.}
\label{fig:h_profile}
\end{figure}

To further investigate the kinematics of the LA members, we explore the Local Standard of Rest radial velocity (V$_{\rm LSR}$) and distance properties of our data in the MS coordinate system. The coordinates are ($\Lambda_{\rm MS}$, $\rm{B_{MS}}$) as defined by \citet{nide08}, where the Magellanic longitude $\Lambda_{\rm MS}$ runs parallel to the MS. We will exclude CD14-A08 as its nature is still unclear (Sec.~\ref{subsubsec:a8}). Fig.~\ref{fig:mage_rv} and Fig.~\ref{fig:mage_dist} show V$_{\rm LSR}$ and distance as functions of ($\Lambda_{\rm MS}$, $\rm{B_{MS}}$), respectively. In Fig.~\ref{fig:mage_rv} we also show the distribution of \ion{H}{1} clouds from \citet{venz12} represented with circles whose diameters increase with the mass of the cloud. As shown in \citet{nide10}, the LA mainly consists of three complexes of gas --  LA I: ($3\degree < \Lambda_{\rm MS} < 29\degree$, $-34\degree < \rm{B_{MS}} < -6\degree$); LA II: ($36\degree < \Lambda_{\rm MS} < 61\degree$, $-17\degree < \rm{B_{MS}} < -10\degree$); and LA III: ($35\degree < \Lambda_{\rm MS} < 62\degree$, $-2\degree < \rm{B_{MS}} < -11\degree$). We mark these areas in our Figs.~\ref{fig:mage_rv} and \ref{fig:mage_dist}. Our kinematical LA members are thus distributed in three different branches of the LA gas: CD14-A05 and CD14-A15 belong to LA~I; CD14-B14 belongs to LA~II; and CD14-B02 and CD14-B03 belong to LA~III. 

From Fig.~\ref{fig:mage_rv}, we can see that the velocities of our five kinematical members (as defined in the previous Section; square symbols) fit well within
the distribution of the \ion{H}{1} clouds of \citet{venz12}. Our LA stars also show a trend of increasing V$_{\rm LSR}$ with $\Lambda_{\rm MS}$. For the relevant coordinate interval ($0\degree < \Lambda_{\rm MS} < 61\degree$, $-35\degree < \rm{B_{MS}} < 5\degree$),  \citet{nide10} find that the V$_{\rm LSR, LA}$ of \ion{H}{1} spreads from $100 - 350$~km~s$^{-1}$ and increases with increasing $\Lambda_{\rm MS}$ (see their Fig. 8). This is in agreement with our findings.

Fig.~\ref{fig:mage_dist} shows that the distances of our kinematically selected LA members range between $12$ and $21$~kpc. These distances correspond to Galactocentric radii between $\sim 11$ and $19$~kpc. \citet{mccl08} derive a kinematic distance to the LA cloud HVC 306-2+230 of 21~kpc, suggesting that the LA crosses the Galactic plane at a Galactic radius of $R \simeq 17$~kpc.  Considering a 20\% distance error estimated by \citet{mccl08}, the distance and the Galactocentric radius of this LA high-velocity cloud is in the range of $17 - 25$~kpc, and $14 - 21$~kpc, respectively. These values agree well with our distance estimates of the LA stars. The scatter we obtain in distance is rather large; perhaps the fact that our stars belong to different LA gaseous regions as marked in Fig.~\ref{fig:mage_dist} may help explain this scatter. There is also a hint that our LA stars below the Galactic plane, and in region LA I have a smaller distance than the LA stars above the Galactic plane. However, this is somewhat uncertain given the two distance estimates of CD14-A05.

\subsection{Final Remarks}
\label{subsec:final_rem}

The average age of the LA members is $59\pm18$~Myr ($67\pm15$~Myr if SA~II parameters are used for CD14-A05). This age is smaller than the LA evolution time scale of $\sim 300$~Myr \citep{diaz12}, suggesting that our LA kinematical members were formed in the LA. The metallicity as represented by Mg in our study indicates that the LA material is more metal poor than the Galactic disk (as represented by two stars at similar location with LA members), and compatible with the metallicity of the LMC (are represented by young cluster NGC 2004). The sole metallicity determination in the LA is that of the gas along the line of sight to Seyfert galaxy NGC 3783 by \citet{lu98}. NGC 3783's sky location is in region LA II. \citet{lu98} find a sulfur abundance of S/H = $0.25\pm0.07$ times solar, which is similar to the SMC metallicity. While, on average our five stars indicate a metallicity compatible with that of the LMC (see \S~\ref{subsec:abund_rv_dist}, if we consider the uncertainty of the [Mg/H] determination of CD14-A05 for instance, we find it is statistically consistent at 90\% confidence with that of cluster NGC 330 in the SMC ($-0.86\pm0.12$). Thus, we cannot exclude the possibility that more metal-poor, SMC-like material could have participated in the formation of CD14-A05 (LA~I) and perhaps CD14-B14 (LA~II). The small size of our sample, combined with the abundance uncertainties, precludes a firm conclusion regarding a more metal-poor, SMC-like origin. 

These results nevertheless confirm the hypothesis that at least parts of the LA are hydrodynamically interacting with the gaseous Galactic disk, forming new stars. These stars are kinematically and chemically distinct from similar-type stars in the Galactic disk. These stars are also relatively close to our Galaxy, compared to the distance to the LMC, i.e., ($\sim 50$ kpc). This confirms that material in the LA was able to reach the edge of the Galactic disk, in agreement with earlier \ion{H}{1} results from \citet{mccl09}.

While we did not find notable LA-kinematic ISM absorption components along the lines of sight to these stars, we did find multi-component features of the ISM in the region below the plane.

\section{Summary}
\label{sec:sum}
Seven element abundances (He, C, N, O, Mg, Si, and S) and kinematics were determined for eight O-/B- type stars in the area of the Magellanic LA, based on high resolution spectra taken with the MIKE instrument on the Magellan 6.5m Clay telescope. 

After investigating the relationship between abundances and kinematics parameters, we found that five stars have kinematics compatible with LA membership, i.e. ${\rm RV} > 100$~km~s$^{-1}$. For the five possible LA member stars,
\begin{enumerate}
\item Mg abundance is significantly lower than that of the remaining two, which are representative of the MW members. Moreover, among the five LA members, four have compatible [Mg/H] with that of B stars in cluster NGC 2004 in the LMC, while [Mg/H] of the remaining one is close to that of cluster NGC 4755 in the MW. Considering the stars' individual uncertainties, we can not statistically exclude the possibility that more metal-poor, SMC-like material could have participated in the formation of CD14-A05 and perhaps CD14-B14. 
\item $V_{\rm LSR}$ decreases with decreasing Magellanic longitude. These decreasing trends are consistent with the conclusions of the LA \ion{H}{1} gas studies of \citet{nide08} and \citet{venz12}. Additionally, the derived distances indicate that the LA is located at a distance of $12 - 21$~kpc, which is in agreement with \ion{H}{1} results from \citet{mccl08, mccl09}.
\item Their average age and small age scatter suggest a single star-forming episode $\sim 60$~Myr ago in the LA.
\end{enumerate}

Our abundance and kinematic results for the LA member stars demonstrate that parts of the LA are hydrodynamically interacting with the gaseous Galactic disk, forming young stars that are chemically distinct from those in the Galactic disk. These results can provide constraints to future models for the Magellanic leading material.

\acknowledgments
We thank Ivan Hubeny for the assistance of model atmosphere and synthetic spectra calculation, Ulrich Heber for discussions of chemical composition of sdO stars, and the anonymous referee for helpful comments. This study was partly supported by National Science Foundation of China (NSFC) grants 11303037 and 11390371/2 and L.Z. acknowledges supports from the Chinese Academy of Sciences (CAS) through a CAS-CONICYT Postdoctoral Fellowship administered by the CAS South America Center for Astronomy (CASSACA) in Santiago, Chile. C.M.B. acknowledges support from project FONDECYT regular 1150060. V.K. acknowledges support from grant N 213/01-216/013-B. R.A.M. acknowledges support from the Chilean Centro de Excelencia en Astrofísica y Tecnologías Afines (CATA) BASAL PFB/06 and from the Project IC120009 Millennium Institute of Astrophysics (MAS) of the Iniciativa Científica Milenio del Ministerio de Economía, Fomento y Turismo de Chile.

\bibliographystyle{apj}
\bibliography{ref}

\clearpage


\clearpage

\clearpage

\clearpage

\clearpage

\clearpage

\clearpage

\clearpage

\clearpage

\clearpage

\clearpage

\clearpage

\clearpage

\clearpage

\clearpage

\clearpage

\clearpage

\clearpage

\clearpage

\begin{deluxetable}{lcccccccc}
\centering
\tabletypesize{\scriptsize}
\tablecolumns{11}
\tablewidth{0pc}
\tablecaption{An observation log from Magellan/MIKE\tablenotemark{a}}
\tablehead{\colhead{ID} & \colhead{SPM ID} &\colhead{R.A.} & \colhead{Dec.} & \colhead{V} & \colhead{Start} & \colhead{Instrument} & \colhead{Exp.}  & \colhead{$S/N$\tablenotemark{b}} \\
                  \colhead{}     & \colhead{}     & \colhead{[J2000]} & \colhead{[J2000]} &   \colhead{}  & \colhead{[UT]} & \colhead{} & \colhead{[s]} & \colhead{} }
\startdata
\multirow{2}{*}{CD14-A05} & \multirow{2}{*}{0390510021} & \multirow{2}{*}{11:10:58.34} & \multirow{2}{*}{-73:14:09.63} & \multirow{2}{*}{15.8} &  \multirow{2}{*}{05:53:54} & MIKE - Blue & \multirow{2}{*}{$1800\times2$} & 19 \\
  &  &  & & & & MIKE - Red &  & 21  \\
\multirow{2}{*}{CD14-A08} &  \multirow{2}{*}{0390226948} & \multirow{2}{*}{11:42:50.38} & \multirow{2}{*}{-76:14:01.06} & \multirow{2}{*}{15.5} &  \multirow{2}{*}{02:54:16} &MIKE - Blue & \multirow{2}{*}{$1900\times3$} & 30  \\
  &  &  & & & & MIKE - Red &  & 32  \\     
\multirow{2}{*}{CD14-A11} & \multirow{2}{*}{0390314266} &  \multirow{2}{*}{11:49:50.47}  &  \multirow{2}{*}{-75:16:23.49} &  \multirow{2}{*}{15.4} &  \multirow{2}{*}{04:18:10} & MIKE - Blue & \multirow{2}{*}{$1800\times2$}  & 18 \\
  &  &  & & & & MIKE - Red &  & 20 \\
\multirow{2}{*}{CD14-A12} & \multirow{2}{*}{0400047074} &  \multirow{2}{*}{11:50:32.16}  &  \multirow{2}{*}{-74:28:37.93} &  \multirow{2}{*}{14.6}  &  \multirow{2}{*}{07:33:20} & MIKE - Blue & \multirow{2}{*}{$1400\times2$} & 25 \\ 
 &  & & & & & MIKE - Red &  & 27  \\
\multirow{2}{*}{CD14-A15} & \multirow{2}{*}{0400042068} &  \multirow{2}{*}{12:19:07.15}  &  \multirow{2}{*}{-74:33:13.61} &  \multirow{2}{*}{15.5}  &  \multirow{2}{*}{09:12:11} & MIKE - Blue & \multirow{2}{*}{$2400\times1 + 2000\times2$} & 31 \\
  &  &  & & & & MIKE - Red &  & 32 \\
\multirow{2}{*}{CD14-B02}   & \multirow{2}{*}{2880084481} &  \multirow{2}{*}{10:14:50.54}  &  \multirow{2}{*}{-43:45:06.30} &  \multirow{2}{*}{13.7}  &  \multirow{2}{*}{00:07:16} & MIKE - Blue & \multirow{2}{*}{$1100\times1$} & 40  \\
  &  &  & & & & MIKE - Red & & 43 \\
\multirow{2}{*}{CD14-B03}   & \multirow{2}{*}{2890097375} &  \multirow{2}{*}{10:33:18.72}  &  \multirow{2}{*}{-42:56:49.19} &  \multirow{2}{*}{14.3}  &  \multirow{2}{*}{00:28:29} & MIKE - Blue & \multirow{2}{*}{$1800\times1$} & 38 \\
 &  &  & & & & MIKE - Red &  & 40 \\
\multirow{2}{*}{CD14-B14} & \multirow{2}{*}{2310104323} &  \multirow{2}{*}{11:27:08.11}  &  \multirow{2}{*}{-48:20:15.18} &  \multirow{2}{*}{14.5}  &  \multirow{2}{*}{01:00:51} & MIKE - Blue & \multirow{2}{*}{$2400\times1$} & 37 \\
  &   & & & & & MIKE - Red &  & 40  \\
\enddata
\label{tab:obs_log}
\tablenotetext{a}{The observation was taken on 2014 March 31.}
\tablenotetext{b}{Signal-to-noise ratios, which were evaluated at 4000 and 5800~{\AA} for blue and red sides, respectively.}
\end{deluxetable}

\clearpage
\begin{deluxetable}{@{\extracolsep{4pt}}lccccc@{}}
\tabletypesize{\scriptsize}
\tablecolumns{6}
\tablewidth{0pc}
\tablecaption{Radial velocity, distance modulus, and age for the target stars.}
\tablehead{\multirow{2}{*}{ID} & \multirow{2}{*}{} & \multirow{2}{*}{RV [km~s$^{-1}$]} & \multirow{2}{*}{age [Myr] }& \multicolumn{2}{c}{$(m-M)_0$} [mag] \\ [1ex]
\cline{5-6} \\ [-1ex]
\colhead{} & \colhead{} & \colhead{} & \colhead{} & \colhead{isochrones} & \colhead{\citet{wegn06}} }
\startdata
\multirow{2}{*}{CD14-A05} & SA~I & \multirow{2}{*}{$133\pm9$} & $25\pm15$ & $15.4\pm0.3$ & $16.3\pm0.2$\\
 & SA~II & & $83\pm7$ & $16.5\pm0.4$ & $..$\\
CD14-A08 & & $48\pm8$& $...$ & $...$ & $...$ \\
CD14-A11 & & $71\pm10$ & $65\pm6$  & $15.7\pm0.5$ & $15.9\pm0.3$ \\
CD14-A12\tablenotemark{a} & & $33\pm7$ & $35\pm5$ & $15.1\pm0.2$ & $15.0\pm0.3$ \\
CD14-A15 & & $162\pm8$ & $70\pm20$ & $15.2\pm0.3$ & $15.6\pm0.2$ \\
CD14-B02 & & $170\pm5$ & $75\pm8$  & $15.3\pm0.2$ & $14.8\pm0.3$ \\
CD14-B03 & & $227\pm6$ & $70\pm10$ & $15.9\pm0.1$ & $15.5\pm0.3$ \\
CD14-B14 & & $202\pm7$ & $70\pm10$ & $16.6\pm0.2$ & $15.6\pm0.3$ \\
\enddata
\label{tab:rv_dis_age}
\tablenotetext{a}{Super-solar metallicity isochrones.}
\end{deluxetable}

\clearpage
\begin{deluxetable}{ccccc}
\tabletypesize{\scriptsize}
\tablecolumns{5}
\tablewidth{0pc}
\tablecaption{The uncertainties of stellar parameters from $S/N$s.}
\tablehead{\multirow{2}{*}{} & \multicolumn{4}{c}{$S/N$s} \\[1ex]
\cline{2-5} \\[-1ex]
\colhead{} & \colhead{10} & \colhead{20} & \colhead{30} & \colhead{40}}
\startdata
$\Delta T_{\rm eff}$~[K]                                                    & 1000    &     800  &   650  & 430\\
$\Delta \log g$~[dex]                                                           &    0.20   &     0.16  &   0.13  & 0.09\\
$\Delta \log \frac{N_{\rm He}}{N_{\rm H}}$      &    0.13    &    0.11  &       0.09  & 0.07\\
\enddata
\label{tab:err_sn} 
\end{deluxetable}

\clearpage
\begin{deluxetable}{@{\extracolsep{4pt}}lcccccccc@{}}
\tabletypesize{\scriptsize}
\tablecolumns{9}
\tablewidth{0pc}
\tablecaption{The stellar parameters measured by two independent methods for target stars. The parameter I are measured by NLTE grids of TLUSTY, and the parameter II re measured by an independent grid (LTE for B stars) and analysis code of LINFOR.}
\tablehead{\multirow{2}{*}{ID} & \multicolumn{2}{c}{$T_{\rm eff}$} & \multicolumn{2}{c}{$\log g$} & \multicolumn{2}{c}{$\log \frac{N_{\rm He}}{N_{\rm H}}$} & \multicolumn{2}{c}{$v\sin i$} \\ [1ex]
 \cline{2-3}  \cline{4-5}   \cline{6-7}  \cline{8-9} \\ [-1ex]
\colhead{} & \colhead{I} & \colhead{II} & \colhead{I} & \colhead{II} & \colhead{I} & \colhead{II} & \colhead{I} & \colhead{II} }
\startdata
CD14-A05   & $15000\pm1000$ & $13700\pm300$     & $4.25\pm0.22$ & $3.81\pm0.09$ & $+0.13\pm0.34$ & $+0.57\pm1.14$ & $110\pm15$ & 130 \\
CD14-A08   & $44250\pm1100$ & $43300\pm1400$  & $4.85\pm0.17$ & $4.29\pm0.12$ & $-1.69\pm0.24$ & $-1.35\pm0.15$   & $25\pm10$ & 50 \\
CD14-A11   & $15750\pm1500$ & $16000\pm900$     & $3.98\pm0.23$ & $3.88\pm0.18$ & $-0.72\pm0.30$ & $-1.29\pm0.18$  & $60\pm10$ & 80 \\
CD14-A12   & $16750\pm1300$ & $16400\pm700$     & $4.00\pm0.20$ & $3.92\pm0.15$ & $-0.07\pm0.26$ & $-0.01\pm0.36$   &  $45\pm10$ & 50 \\
CD14-A15   & $14750\pm1200$ & $14500\pm500$     & $4.10\pm0.20$ & $4.11\pm0.15$ & $-0.65\pm0.25$ & $-1.55\pm0.18$  & $230\pm20$ & 230 \\
CD14-B02   & $15100\pm600$    & $15000\pm300$     & $3.70\pm0.12$ & $3.65\pm0.09$ & $-0.57\pm0.18$ & $-0.55\pm0.12$  & $22\pm5$ & 30 \\
CD14-B03   & $15650\pm700$    & $15500\pm500$     & $3.71\pm0.13$ & $3.68\pm0.09$ & $-1.14\pm0.19$ & $-1.30\pm0.12$  & $15\pm10$ & 0 \\
CD14-B14   & $15250\pm1000$   & $15400\pm700$      & $3.50\pm0.21$ & $3.29\pm0.15$ & $-0.70\pm0.25$ & $-0.75\pm0.24$  & $310\pm20$ & 310 \\
\enddata
\label{tab:st_params}
\end{deluxetable}

\clearpage

\begin{deluxetable}{crrrr}
\tabletypesize{\scriptsize}
\tablecolumns{5}
\tablewidth{0pc}
\tablecaption{Comparisons of stellar parameters between the two analyses.}
\tablehead{\colhead{ID} & \colhead{$\Delta T_{\rm eff}$} & \colhead{$\Delta \log g$} & \colhead{$\Delta \log \frac{N_{\rm He}}{N_{\rm H}}$} & \colhead{$\Delta v \sin i$}}
\startdata
CD14-A05   & $1300\pm 1044$    & $0.44 \pm 0.24$  & $-0.44 \pm 1.19$  & $-20 \pm 15$\\
CD14-A08   & $950 \pm 1780$    & $0.56 \pm 0.21$  & $-0.34 \pm 0.28$  & $-25 \pm 10$\\
CD14-A11   & $-250 \pm 1655$   & $0.10 \pm 0.29$  & $0.57 \pm 0.35$   & $-20 \pm 10$\\
CD14-A12   & $350 \pm 1390$    & $0.08 \pm 0.25$  & $-0.06 \pm 0.44$  & $-5 \pm 10$\\
CD14-A15   & $250 \pm 1240$    & $-0.01 \pm 0.25$ & $0.90 \pm 0.31$   & $0 \pm 20$ \\
CD14-B02   & $100 \pm 780 $    & $0.05 \pm 0.15$  & $-0.02 \pm 0.21$  & $-8 \pm 5$\\
CD14-B03   & $150 \pm 990$     & $0.03 \pm 0.16$  & $0.16 \pm 0.22$   & $15 \pm 10$\\
CD14-B14   & $-150 \pm 1410$   & $0.21 \pm 0.26$  & $0.05 \pm 0.35$   & $0 \pm 20$ \\
\enddata
\label{tab:sp_com}
\tablecomments{$\Delta$ = SA~I - SA~II}
\end{deluxetable}

\clearpage
\begin{deluxetable}{llcccc}
\tabletypesize{\scriptsize} 
\tablecolumns{6}
\tablewidth{0pc}
\tablecaption{Abundance uncertainties linked to stellar parameters. } 
\tablehead{Star & Abundance & $\Delta T_{\rm eff}$ ($\pm1100$~K) & $\Delta \log g$ ($\pm0.17$) & $v\sin i$ ($\pm10$~km~s$^{-1}$) &Total Error } 
\startdata
CD14-A08             & $\log \epsilon ({\rm He})$ & $\pm0.15$ & $\mp0.14$ & $\pm0.08$ & $\pm0.22$  \\
O-type star           & $\log \epsilon ({\rm C})$   & $...$ & $...$ & $...$  & $...$   \\
($<S/N> = 26$)  & $\log \epsilon ({\rm N})$   & $\pm0.16$ & $\pm0.15$ & $\pm0.09$ & $\pm0.24$  \\
                                     & $\log \epsilon ({\rm O})$   & $\mp0.13$ & $\pm0.14$ & $\pm0.10$ & $\pm0.22$  \\
                                    & $\log \epsilon ({\rm Mg})$ & $\pm0.14$ & $\pm0.15$ & $\pm0.07$& $\pm0.21$  \\
                                    & $\log \epsilon ({\rm Si})$  &  $...$ & $...$ & $...$  & $...$   \\
                                    & $\log \epsilon ({\rm S})$   &  $...$ & $...$ & $...$  & $...$   \\
\tableline
 & &$\Delta T_{\rm eff}$ ($\pm600$~K) &  $\Delta \log g$ ($\pm0.12$) &  $v\sin i$ ($\pm 5$~km~s$^{-1}$) & \\
\tableline           
 CD14-B02          & $\log \epsilon ({\rm He})$ & $\pm0.10$ & $\mp0.08$ & $\pm0.07$ & $\pm0.15$  \\
B-type star           & $\log \epsilon ({\rm C})$  & $\pm0.10$ & $\pm0.08$ & $\pm0.06$ & $\pm0.14$  \\
($<S/N> = 37$) & $\log \epsilon ({\rm N})$  & $\pm0.09$ & $\pm0.08$ & $\pm0.06$ & $\pm0.13$  \\
                                    & $\log \epsilon ({\rm O})$  & $\mp0.10$ & $\pm0.06$ & $\pm0.07$ & $\pm0.14$  \\
                                   & $\log \epsilon ({\rm Mg})$ & $\pm0.10$ & $\pm0.07$ & $\pm0.07$ & $\pm0.14$  \\
                                   & $\log \epsilon ({\rm Si})$ & $\pm0.11$ & $\pm0.08$ & $\pm0.07$ & $\pm0.15$  \\
                                   & $\log \epsilon ({\rm S})$  & $\mp0.11$ & $\pm0.08$ & $\pm0.06$ & $\pm0.15$  \\
\enddata
\label{tab:err_at_pro}
\end{deluxetable}

\clearpage
\begin{deluxetable}{lcccccccc}
\tabletypesize{\scriptsize} 
\tablecolumns{9}
\tablewidth{0pc}
\tablecaption{Element abundance results of the targets} 
\tablehead{ & CD14-A05 & CD14-A08 & CD14-A11  & CD14-A12  & CD14-A15 & CD14-B02 & CD14-B03 & CD14-B14} 
\startdata
$\log \epsilon ({\rm He})$ & $12.06\pm0.34$ & $10.24\pm0.24$ & $11.21\pm0.30$ & $11.86\pm0.26$ & $11.28\pm0.25$   & $11.36\pm0.18$ &  $10.79\pm0.19$ & $11.23\pm0.25$ \\                       
$\log \epsilon ({\rm C})$   & $  9.20\pm0.32$   & $<8.02$                   & $  8.41\pm0.24$   & $ 9.71\pm0.25$   &  $<9.25$                     &  $9.09\pm0.16$    &  $  8.58\pm0.18$ & $< 8.82$            \\            
$\log \epsilon ({\rm N})$   & $...$                              & $8.31\pm0.26$   & $...$                               & $< 9.45$                    &  $...$                               &   $ 8.17\pm0.25$  &  $8.89\pm0.25$  &  $...$                   \\            
$\log \epsilon ({\rm O})$   & $...$                              & $8.72\pm0.31$   & $...$                               & $<9.79$                     &  $...$                               &   $9.49\pm0.20$   &  $9.61\pm0.20$   &  $...$                   \\        
$\log \epsilon ({\rm Mg})$ & $ 7.01\pm0.35$ & $8.19\pm0.23$    & $7.46\pm0.27$   & $7.57\pm0.25$      &  $7.11\pm0.24$      &  $7.48\pm0.16$  &  $7.14\pm0.18$   &  $7.08\pm0.29$  \\     
$\log \epsilon ({\rm Si})$  &  $< 8.25$                  & $< 7.42 $                  &  $7.45\pm0.29$   & $<7.75$                      &  $< 8.05$                     &  $7.57\pm0.16$  & $7.36\pm0.18$   &  $<8.05$                   \\     
$\log \epsilon ({\rm S})$   &  $<8.50$                   & $...$                              & $<8.33$                     & $<7.93$                     &  $...$                                &  $7.63\pm0.23$  & $7.71\pm0.25$   &   $...$                   \\      
\hline
${\rm [He/H]}$ \tablenotemark{a} & $1.13\pm0.34$      & $-0.69\pm0.24$  & $  0.28\pm0.30$  & $ 0.93\pm0.26$  & $ 0.35\pm0.25$   & $ 0.43\pm0.18$ &  $-0.14\pm0.19$ & $0.30\pm0.25$   \\
${\rm [C/H]}$                                              & $ 0.68\pm0.32$     & $< -0.51$                  & $-0.11\pm0.24$  & $ 1.19\pm0.25$  &  $<0.73$                   & $ 0.57\pm0.16$ &  $0.05\pm0.18$  & $< 0.30$           \\            
${\rm [N/H]}$                                              & $...$                                & $0.39\pm0.26$    & $...$                              & $< 1.53$                  &  $...$                             & $0.25\pm0.25$  &  $ 0.97\pm0.25$ &  $...$                   \\            
${\rm [O/H]}$                                              & $...$                                & $0.11\pm0.26$  & $...$                              & $< 0.96$                  &  $...$                             &  $0.66\pm0.20$  &  $0.78\pm0.20$  &  $...$                   \\        
${\rm [Mg/H]}$                                          & $ -0.57\pm0.35$    & $0.61\pm0.23$   & $ -0.12\pm0.27$ & $-0.01\pm0.25$ &  $-0.47\pm0.24$  &  $-0.10\pm0.16$  &  $-0.44\pm0.18$ &  $-0.50\pm0.29$  \\     
${\rm [Si/H]}$                                             &  $< 0.70$                     & $< -0.13$                 &  $-0.10\pm0.29$ & $<0.20$                   &  $< 0.5$                      &  $0.02\pm0.16$  & $-0.19\pm0.18$  &  $<0.5$                   \\     
${\rm [S/H]}$                                              &  $< 1.00$                      & $...$                             & $<1.00$                     & $<0.60$                   &  $...$                              &  $0.30\pm0.23$  & $ 0.38\pm0.25$  &   $...$                   \\      
\enddata
\label{tab:abun_results_a}
\tablenotetext{a}{Solar compositions are taken from \citet{grev98}}
\end{deluxetable}

\end{document}